\documentclass[useAMS,usenatbib]{mnras}

\usepackage{graphicx}
\usepackage[T1]{fontenc}
\usepackage{aecompl}
\usepackage[dvipsnames]{color}
\usepackage[normalem]{ulem}
\usepackage{threeparttable}
\usepackage{multirow}
\usepackage{array}

\newcommand{\SII}{[S~{\sc ii}]\ }
\newcommand{\OIII}{[O~{\sc iii}]\ }
\newcommand{\NII}{[N~{\sc ii}]\ }
\newcommand{\HII}{H~{\sc ii}\ }
\newcommand{\HeII}{He~{\sc ii}\ }
\newcommand{\HI}{H~{\sc i}\ }
\newcommand{\Ha}{H$\alpha$\ }
\newcommand{\kms}{\,\mbox{km}\,\mbox{s}^{-1}}

\newcommand{\HST}{\textit{HST}\ }
\newcommand{\SIIHa}{I([S~{\sc ii}])/I(H$\alpha$)}
\newcommand{\SIIOIII}{I([S~{\sc ii}])/I([O~{\sc iii}])}
\newcommand{\NIIHa}{I([N~{\sc ii}])/I(H$\alpha$)}
\newcommand{\OIIIHb}{I([O~{\sc iii}]5007)/I(H$\beta$)}
\newcommand{\be}{\begin{equation}}
\newcommand{\ee}{\end{equation}}

\def \gtsima{$\, \buildrel > \over \sim \,$}
\def \ltsima{$\, \buildrel < \over \sim \,$}
\def \simgt{\lower.5ex\hbox{\gtsima}}
\def \simlt{\lower.5ex\hbox{\ltsima}}

\begin{document}

\title[Complexes of SF in SGS of Holmberg~II.] {Complexes of triggered star formation
in supergiant shell of Holmberg~II.}

\author[Egorov et al.]{
    Oleg V.~Egorov$^{1}$\thanks{E-mail: egorov@sai.msu.ru},
    Tatiana A.~Lozinskaya$^{1}$,
    Alexei V.~Moiseev$^{1,2}$,
    and Yuri A.~Shchekinov$^{3}$ \\
    $^{1}$ Lomonosov Moscow State University, Sternberg Astronomical Institute,
    Universitetsky pr. 13, Moscow 119234, Russia
    \\
    $^{2}$ Special Astrophysical Observatory, Russian Academy of Sciences, Nizhny Arkhyz 369167, Russia
    \\
    $^{3}$ P. N. Lebedev Physical Institute, 53 Leninskiy Prospekt, 119991 Moscow, Russia
}

\date{Accepted 2016 Month 00. Received 2016 Month 00; in original
    form 2016 Month 00}

\pagerange{\pageref{firstpage}--\pageref{lastpage}} \pubyear{2016}

\maketitle

\label{firstpage}

\begin{abstract}

We report a detailed analysis of all regions of current star formation in the walls of the supergiant \HI shell (SGS) in the galaxy Holmberg~II based on observations with a scanning Fabry--Perot interferometer at the 6-m SAO RAS telescope. We compare the structure and kinematics of ionized gas with that of atomic hydrogen and with  the stellar population of the SGS. Our deep \Ha images and archival images taken by the \textit{HST} demonstrate that current star formation episodes are larger and more complicated than previously thought: they represent unified star-forming complexes with sizes of several hundred pc rather than `chains' of separate bright nebulae in the walls of the SGS. The fact that we are dealing with  unified complexes is evidenced by identified faint shell-like structures of ionized and neutral gas which connect several distinct bright \HII regions. Formation of such complexes is due to the feedback of stars with very inhomogeneous ambient gas in the walls of the SGS. The arguments supporting an idea about the triggering of star formation in SGS by the \HI supershells collision are presented. We also found a faint ionized supershell inside the \HI SGS expanding with a velocity of no greater than $10-15 \kms$. Five OB stars located inside the inner supershell  are sufficient to account for its radiation, although a possibility of leakage of ionizing photons from bright \HII regions is not ruled
out as well.

\end{abstract}


\section{Introduction}

\begin{figure}
   \includegraphics[width=\linewidth]{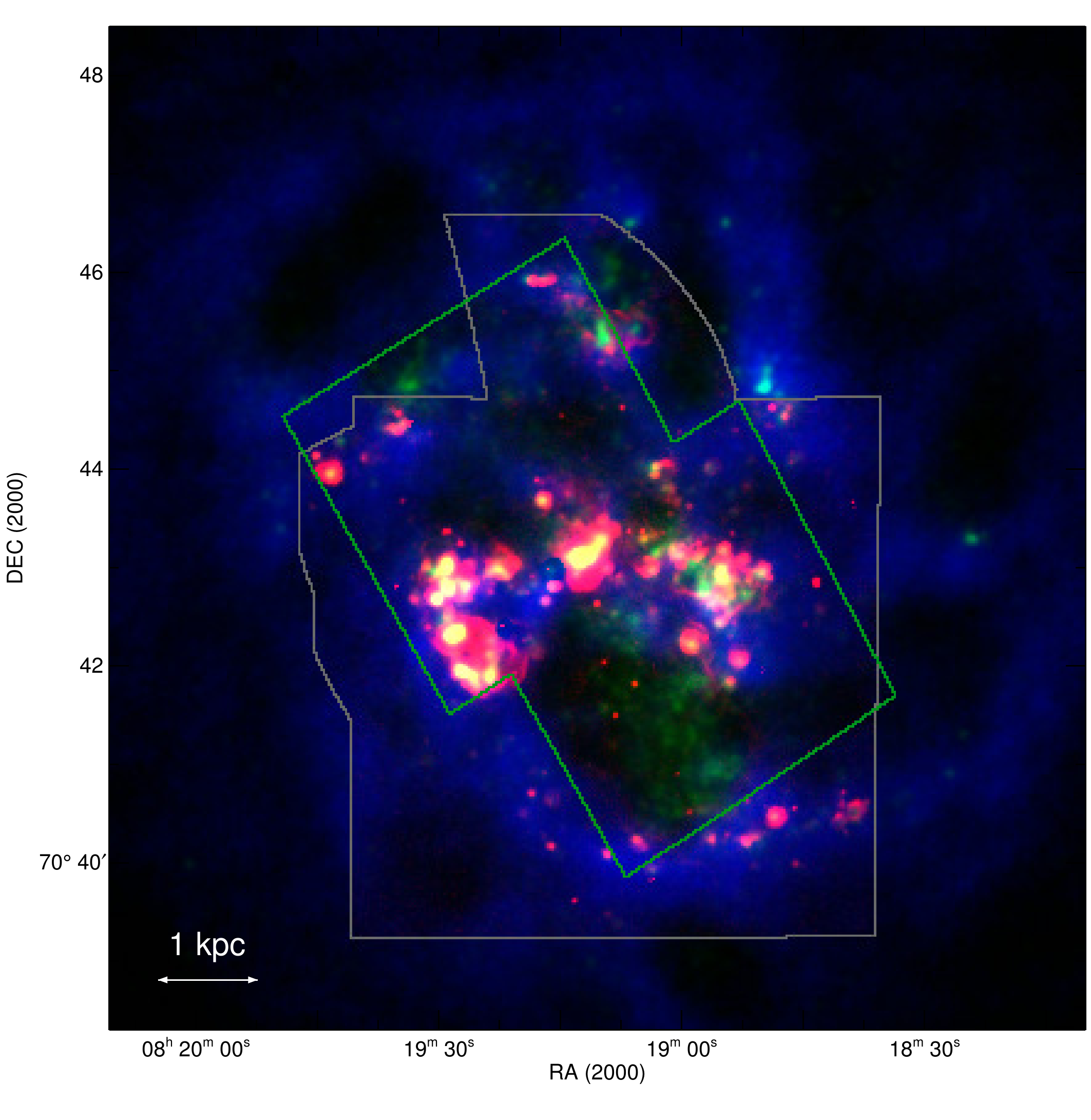}

      \includegraphics[width=\linewidth]{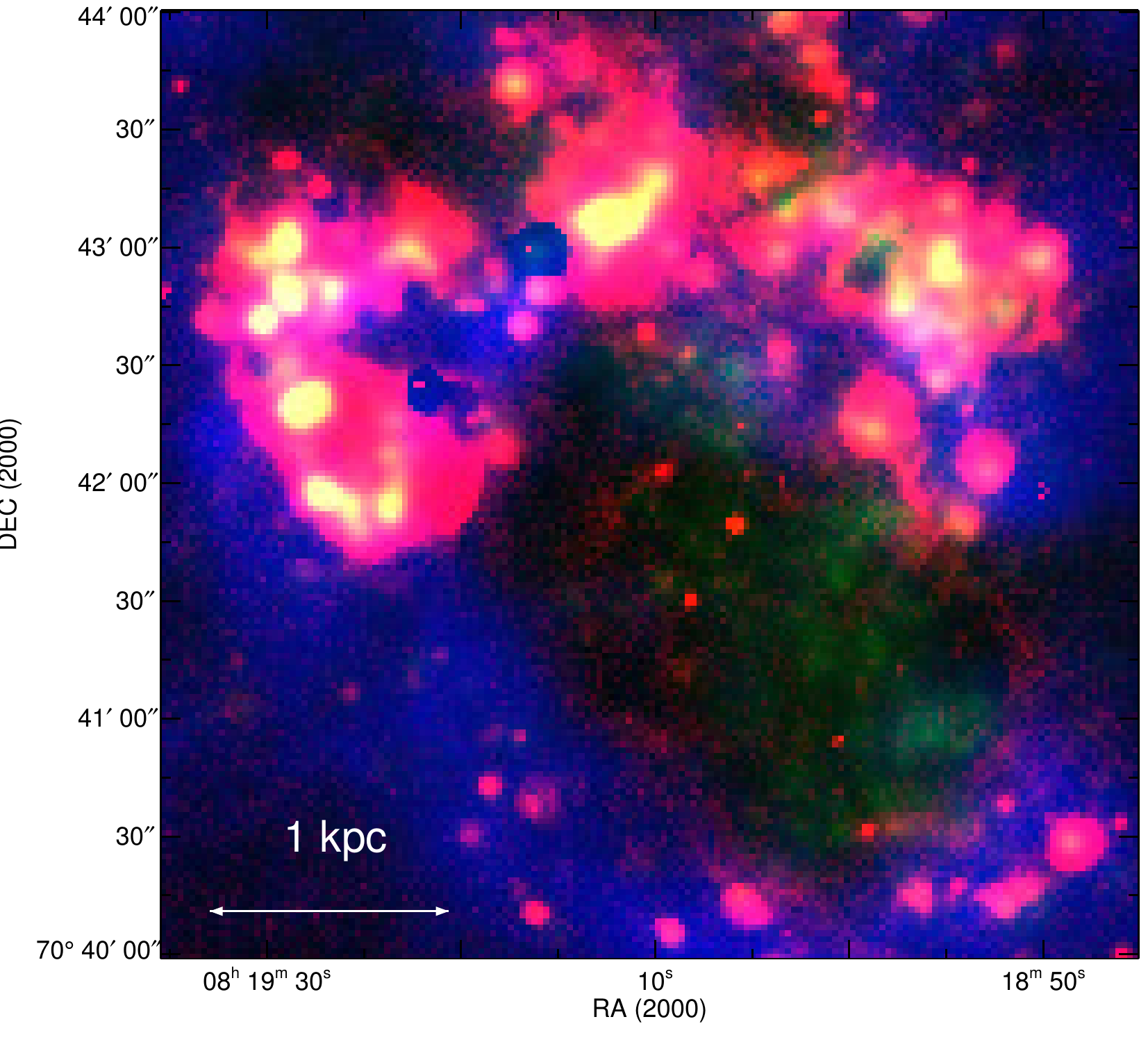}
    \caption{False-colour image of the galaxy Holmberg~II: red, green, and blue channels correspond to H$\alpha$,  FUV (GALEX) and  \HI 21 cm (LITTLE THINGS) emission, respectively.
        The grey line denotes  the area covered by our \Ha observations 
        (both imaging and FPI) excluding the regions with optical ghosts contamination;
        the green line shows the area covered by \HST data we used to identify
        OB-stars. Bottom image shows the area of active star formation
        with the different brightness setup}\label{fig:colorIm}
\end{figure}

Irregular galaxies are widely used to study stellar feedback: radiation, stellar winds
and supernovae that regulate the structure and kinematics of the interstellar medium
(ISM) and might trigger new episodes of star formation. Due to the lack of spiral waves and the fact that gaseous discs are thicker than those in spiral galaxies, irregular galaxies reveal giant \HI supershells and holes with sizes as large as 1 -- 2 kpc and lifetimes up to several hundreds Myr.   Giant supershells and holes in some galaxies represent
    the dominant feature of the ISM  \citep*[see, e.g.][and references therein]{Young97, ott01, Simpson05, cannon11a, Warren11}. Such large structures are usually called supergiant shells (SGS), or giant supershells.

Formation mechanisms of supergiant shells have been discussed extensively in recent decades. In the standard approach based on the \citet{weaver77} model, the cumulative action of multiple stellar winds and supernovae
explosions is responsible for \HI shells formation  \citep[see, e.g.][]{mccray87, tenor88, ott01}. However, it has been recognized long ago \citep[see, e.g.][]{tenor88, rhode99, kim99,  Simpson05, silich06} that this scenario cannot explain the origin of giant supershells, in which the mechanical energy input from detected stellar clusters appears to be inconsistent with that required by the standard  model. Several other mechanisms 
have been proposed: collisions of high velocity clouds with galactic
    discs \citep{tenor81}, fractal ISM \citep{elmegreen97}, radiation pressure from field
    stars \citep{elmegreen82}, ram pressure of the intergalactic medium
    \citep{bureau02}, \HI dissolution by UV radiation \citep{vorobyov04}, non-linear evolution of
    self-gravitating turbulent galactic discs \citep*{wada00, dib05} and even exotic mechanisms (see references in \citealt{silich06} and \citealt{Warren11}).

The Im type (Magellanic-type 
irregulars) non-interacting gas-rich galaxy Holmberg~II,  a
member of the  M81 -- NGC~2403 group, is one of the main
`laboratories' for studying this phenomenon. \citet{puche92}
explained the formation of SGSs and giant \HI cavities in
Holmberg~II by the effect of multiple supernova explosions and
stellar winds of massive stars located inside the associations.
\citet{maschenko95} adopted this mechanism of SGSs  formation and
explained the discrepancy between the computed and observed
elongation  of many such shells by their mutual interaction and
inhomogeneity of the ambient medium. The problem of the
validity of such a `standard mechanism' in the case of
Holmberg~II was first considered by \citet{rhode99} and then
discussed by \citet{bureau02} and \citet{weisz09a}. The trouble
with it was that many supershells lacked young stellar
associations whose energy could be sufficient to produce the
observed structures.

By now, it is well established that high star-formation rates can
be maintained for a much longer time period than previously
believed: dwarf galaxies are characterized by the ability to
sustain a high star-formation efficiency over several hundred
Myr, and  in some cases even up to 1~Gyr -- the specific (per stellar mass
unit) star formation rate is high enough to exhaust
gas content in $0.4-1$ Gyr  \citep{mcquinn09,
    mcquinn10a, mcquinn10b} with local short starbursts occurring
during this period. Such long periods of intense star formation
provide enough energy from stellar winds and supernova explosions
to drive the formation of giant cavities and supershells
\citep[see, e.g.][]{weisz09b, cannon11a, cannon11b, Warren11}.
\citet{weisz09a} have shown
that the supershells and holes in the galaxy Holmberg~II contain multiple
stellar generations and might have been formed from the energy
input of these stars integrated over the lifetime of the \HI
structure. A comparison of the energy of  supergiant shells with
the energy inflow from supernovae performed by
\citet{bagetakos11} is also consistent with their formation as a
result of multiple starbursts spanning a long period of time.

The energetics of supergiant shells, i.e. the fraction of energy deposited in them by supernovae explosions and stellar winds, is a crucial issue for understanding their origin. It is commonly known since the first numerical studies of dynamics of supernovae remnants \citet{cox72,chevalier74} that after entering the radiative expansion stage, the total energy of the remnant decreases with a radius of $\propto R^{-2}$ to $\propto R^{-3}$, so that on a time-scale of 0.1 Myr, only around 2\% of the SN energy is available for driving the shell \citep[see more recent simulations with non-equilibrium cooling by][]{sharma14}. \citet{tomisaka81} assumed that the fraction of energy radiatively lost by a supernova might decrease, when an explosion occurs into a low density bubble produced by a stellar wind and/or previously exploded supernova(e). Within a simplistic model of sequential explosions of 20 to 100 SNe occurring at the same centre, they found qualitatively reasonable agreement with galactic superbubbles. 
The principal issues: the exact expansion law and the fraction of energy remained for a continuous support of growing superbubbles, have been nonetheless left beyond clear understanding. 
In full 3D numerical simulations of sequential SNe exploding at the same centre, \citet{sharma14} have found that even on a long time-scale (30 to 50 Myr), up to 30\% of the injected  energy can retain into slow expansion of the supershell. This conclusion is valid though only in case of `coherent' SNe explosions, i.e. when the remnants overlap before entering the radiative stage \citep{nath13}. Numerical models with scattered explosions, i.e. spread through over an active volume, confirm this result: collective action of `coherent' SNe makes them more efficient in driving supershells, with a fraction of retained energy of $\sim 0.1$ on times up to 10 Myr \citep{vasiliev15,yadav16}. In all cases, the expansion sets asymptotically into a power law close to the standard $t^{3/5}$ wind regime with mechanical luminosity reduced by a factor of approximately 10, though still depending on interrelation between gas density and SN rate \citet{vasiliev15}. When the origin of supergiant shells of kpc-scales are concerned, particularly in such galaxies as Ho~II and IC 2574, where signs of merging of different superbubbles are clearly seen, numerical description becomes more challenging. However, in conditions typical of dwarf galaxies, the wind expansion law with energy reduced by a factor of 10 would be a reasonable conservative approach.

Short (about 10 Myr) local starbursts, which we observed as
complexes of ionized gas, are located mainly inside dense walls
of giant  H~{\sc i} structures in Holmberg~II. According to the modern concepts,
ongoing star formation there could be triggered by the expansion
and/or collision of supergiant shells. The analysis of the
influence of these new sites of star formation on the evolution
of supergiant shells and hence on the structure of the gaseous
medium of the galaxy is of greatest interest.


\medskip

This work is a third in the series of our investigations of the galaxy Holmberg~II. Previously in \citet*{egorov13},
we studied the ionized gas spectra of star-forming regions in
Ho~II  using optical long-slit spectroscopic observations carried out
with the 6-m SAO RAS telescope. We estimated oxygen,
nitrogen, sulphur, neon and argon abundances in individual \HII
regions and found the average metallicity in the galaxy to be
either 0.1 $Z_\odot$ or 0.3 $Z_\odot$ depending on the
estimation method applied.

In \citet{wiebe14},
we performed a multiwavelength photometric study of star formation regions in Ho~II using archival data of GALEX, \textit{Spitzer} and \textit{Herschel}
space telescopes. We have examined for the first time how the emission of star-forming regions over a wide
wavelength range evolves with time.  We traced the evolution of the fraction of polycyclic aromatic
hydrocarbons (PAHs) and of the hot-grain properties in the galaxy.

The aim of this study is to analyse the kinematics of ionized gas in all regions of star formation and of their
ambient neutral gas in the most extended 2~kpc sized supergiant shell in the galaxy.
We search for shell-like structures of ionized gas that reveal
signs of expansion and compare them with stellar population
in star-forming regions. We also try to recognize kinematic
evidences for possible supernovae shocks  in the eastern chain of
bright emission nebulae predicted by \citet{tong95}.

An analysis of gas kinematics 
in the rim of an  \HI SGS is of great interest, because it may help
to find out how such giant structures change when affected by new
local bursts of triggered star formation. When studying the
supergiant  \HI shell with triggered star formation located in
the  Irr galaxy IC~2574 \citep{egorov14}, we found for the first
time an inner  weak diffuse  emission of ionized gas in \Ha and
\SII 6717,6731 lines with kinematic evidences for expansion. We
showed that such an inner giant ionized supershell must have likely be
formed as a result of the action of Lyman continuum photons escaping the
star-forming regions on to the neutral gas in the supergiant \HI
shell. This is a new phenomenon hitherto unobserved in  IC~2574,
and the search of such inner shell-like
\HII structures inside \HI supergiant shells in other irregular
galaxies seems to be interesting. Earlier, we briefly announced the discovery of a similar giant  faint  ionized supershell inside
the \HI SGS in Holmberg~II   \citep*{egorov15}, which is seen on the bottom panel of Fig.~\ref{fig:colorIm}; here we report the results of its  investigation.

To avoid confusion, further we will use the SGS
abbreviation to refer to the supergiant shell in Holmberg~II under consideration
and will use the full term `supergiant shell' to denote the entire
class of such objects.

This study is based on our observations conducted with the scanning
Fabry--Perot  interferometer  (FPI) in \Ha line and with narrow-band
\Ha, \SII and \OIII filters at the 6-m telescope of the Special
Astrophysical Observatory of the Russian Academy of Sciences (SAO
RAS). High spatial and spectral resolutions of FPI data allows to analyze the ionized gas kinematics in superbubbles in details. A number of the studies made using this technique were published \citep*[see e.g.][and references therein]{lozinsk03, relano05, egorov10, egorov14, cf15, sc15}

The observations and data reduction performed are described
in Section~\ref{sec:obs}. In Section~\ref{sec:overview}, we
overview the results of previous multiwavelength studies of Holmberg~II.
Section~\ref{sec:shells} is devoted to the search for expanding
ionized and neutral shells and supershells. In
Section~\ref{sec:morphology}, we analyse the morphology of \HII
complexes in the galaxy. In Section~\ref{sec:discussion}, we
discuss the results obtained, including the nature of the faint
ionized inner supershell inside the \HI SGS.
Section~\ref{sec:summary} summarizes the main results of this
study.

\section{Observations and data reduction}\label{sec:obs}

\subsection{Optical FPI observations}

\begin{table*}
    \caption{Log of the observed data}
    \label{tab:obs_data}
    \begin{tabular}{llllllll}
        \hline
        Data set            & Date of obs & $\mathrm{T_{exp}}$, s  & FOV &  $''/px$  & $\theta$, $''$ &  sp. range  & $\delta\lambda$  ($\delta v$) \\
        \hline
        FPI field \#1& 07/08 Feb 2010 & $40\times150$  &   {$6.1'\times6.1'$} & {0.71}  &  1.8 & {8.8~\AA\, around \Ha} & {0.48~\AA \,($22\kms$) } \\
        FPI field \#2 & 26/27 Apr 2011 & $40\times240$  &   $6.1'\times6.1'$  &  0.71   & 1.4  & {8.8~\AA\, around \Ha} & {0.48~\AA \,($22\kms$) } \\
        FPI field \#3 & 16/17 Dec 2014 & $40\times160$   &  $6.1'\times6.1'$   & 0.71    & 1.4  & {8.8~\AA\, around \Ha} & {0.48~\AA \,($22\kms$) } \\
        FN655 image        & 18/19 Oct 2014 & 3600                   & $6.1'\times6.1'$                       & 0.36                      &  1.9 &    \Ha + \NII & \\
        FN674 image        & 18/19 Oct 2014 & 1800                   & $6.1'\times6.1'$                       & 0.36                      &  2.0 &     \SII 6717,6731~\AA & \\
        IFP502 image        & 02/03 Mar 2016 & 2400                   & $6.1'\times6.1'$                       & 0.36                      &  1.8 &     \OIII 5007~\AA & \\
        FN641 image        & 18/19 Oct 2014 & 1200                   & $6.1'\times6.1'$                       & 0.36                      &  2.0 &     Continuum & \\
        FN712 image        & 18/19 Oct 2014 & 1200                   & $6.1'\times6.1'$                       & 0.36                      &  2.0 &     Continuum & \\
        SED525 image        & 02/03 Mar 2016 & 1200                   & $6.1'\times6.1'$                       & 0.36                      &  2.0 &     Continuum & \\
        \hline
    \end{tabular}
\end{table*}

\

The observations were carried out at the prime focus of the 6-m SAO RAS
telescope using a scanning  FPI mounted  inside the
SCORPIO-2  multi-mode focal reducer \citep{scorpio2}. The operating spectral range
around the \Ha emission line was cut by a narrow bandpass filter with a $\mathrm{FWHM}\approx14$~\AA\
bandwidth. The interferometer
provides a free spectral range between the neighbouring interference orders $\Delta\lambda=8.8$~\AA\, with
a spectral resolution (FWHM of the instrumental profile) of about $0.48$~\AA. During the scanning process,
we have consecutively obtained 40 interferograms at different distances
between the FPI plates. The log of these observations and parameters of the other data sets are
given in Table~\ref{tab:obs_data}, where $\mathrm{T_{exp}}$ is the exposure time; FOV -- the field of
view;  $''/px$ -- the  pixel size; $\theta$ -- the final angular resolution;  $\delta\lambda$ and $\delta v$
are the final spectral and velocity resolution.

The data reduction was performed using a software package running in the \textsc{idl} environment.
For a detailed description of the data reduction algorithms,
see \citet{Moiseev02ifp, MoiseevEgorov2008, Moiseev2015}.
After the initial reduction, sky line subtraction, photometric and seeing corrections made using the
reference stars, and wavelength calibration, the observational data were combined into data cubes,
where each pixel in the field of view contains a 40-channel spectrum. The calibration wavelength was 6598.95 \AA.
We observed the galaxy in three overlapped fields (see Table~\ref{tab:obs_data}), each field was exposed at two
position angles in order to remove the parasitic ghost reflection. The regions of interest are filled less then a half of each observed field. This technique allowed us to get rid of the ghost contamination almost everywhere at the cost of reduced signal-to-noise ratio in the area of the ghosts (see details of this algorithm in \citealt{MoiseevEgorov2008}). The remaining symmetric ghosts are located in the outer regions of the total observed field. All the central regions studied in this paper are free from ghost contamination.

The data for all fields were reduced separately to obtain the wavelength cubes of the object.
After the data reduction, we constructed mosaic of these three  fields. We found the data for field \#1
to be of poor quality (variable atmospheric conditions) compared to the data for the two other fields.
That is why we included it in the mosaic only in the area of the faint inner supershell
(see Section~\ref{sec:superbubble}) in order to increase the signal-to-noise ratio there
(the location of Field \#1 coincides with that of~\#3).

In order to get rid of the stellar continuum emission in the data cube, we performed a zero-order background
fit for each pixel at the edges of line profiles and then subtracted it.  The analysis of  \Ha line profiles
was carried out using the multi-component Voigt fitting  \citep{MoiseevEgorov2008}.

The gas morphology and kinematics in the region of the interaction of multiple supershells  are very
complex and it would therefore be useful to exclude the regular component associated with the rotation of
the galaxy to understand its local kinematics. We constructed a rotation model of Ho~II from the \HI
velocity field using the tilted-ring approximation adapted to dwarf galaxy
kinematics \citep{Moiseev2014} with initial parameters for Ho~II taken from \citet{Oh11} (we list them in
Table~\ref{tab:main_params}) and then subtracted this model from the observed data cube.
As a result, we have obtained the data cube that reflects the local motions caused by stellar feedback.
In the final cube, zero velocity corresponds to the regions that show no signs of the feedback influence
on the kinematics of ionized gas. We described this `derotation' method in \citet{egorov14}.

\begin{figure}
    \includegraphics[width=\linewidth]{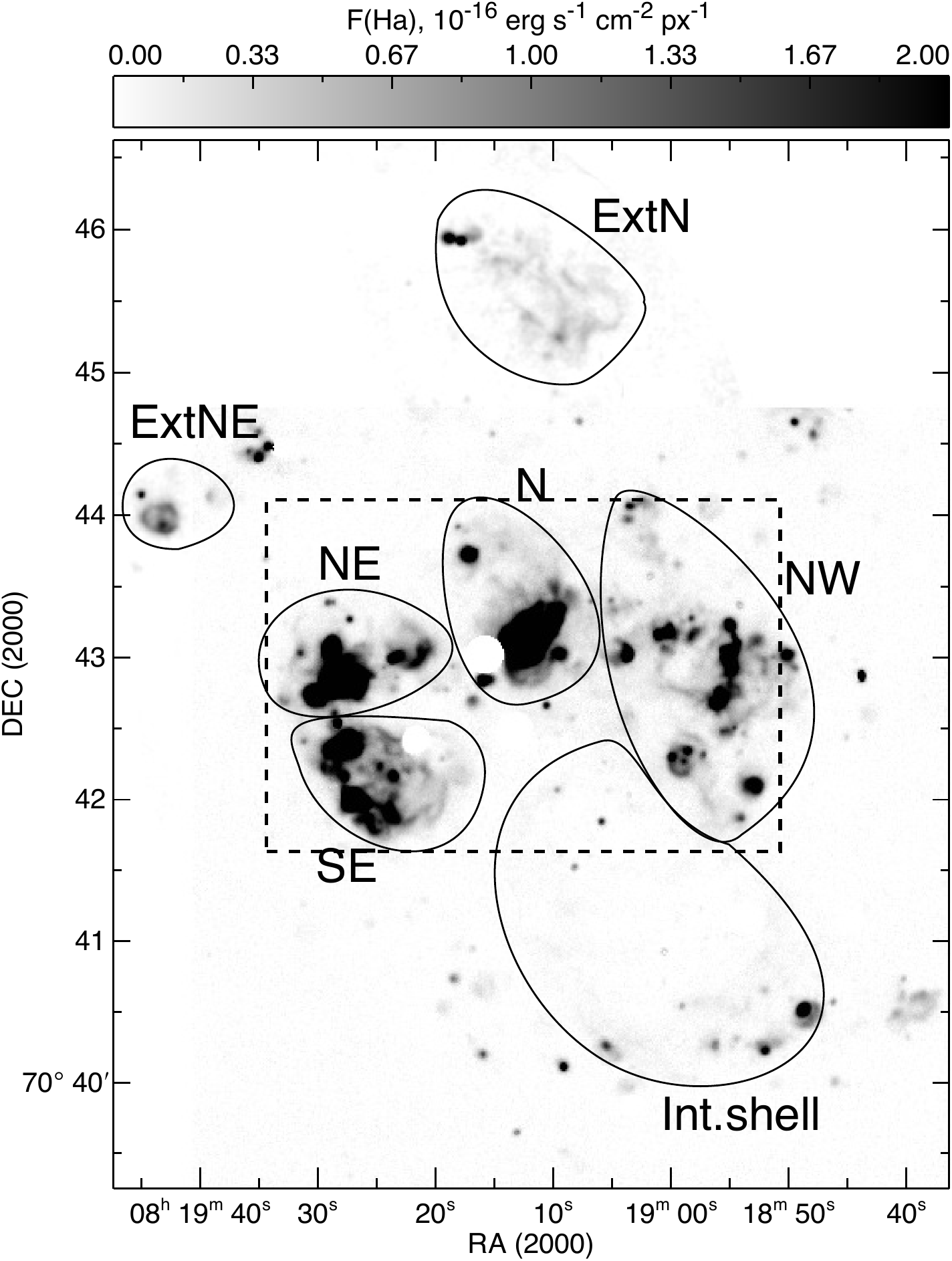}

    \includegraphics[width=\linewidth]{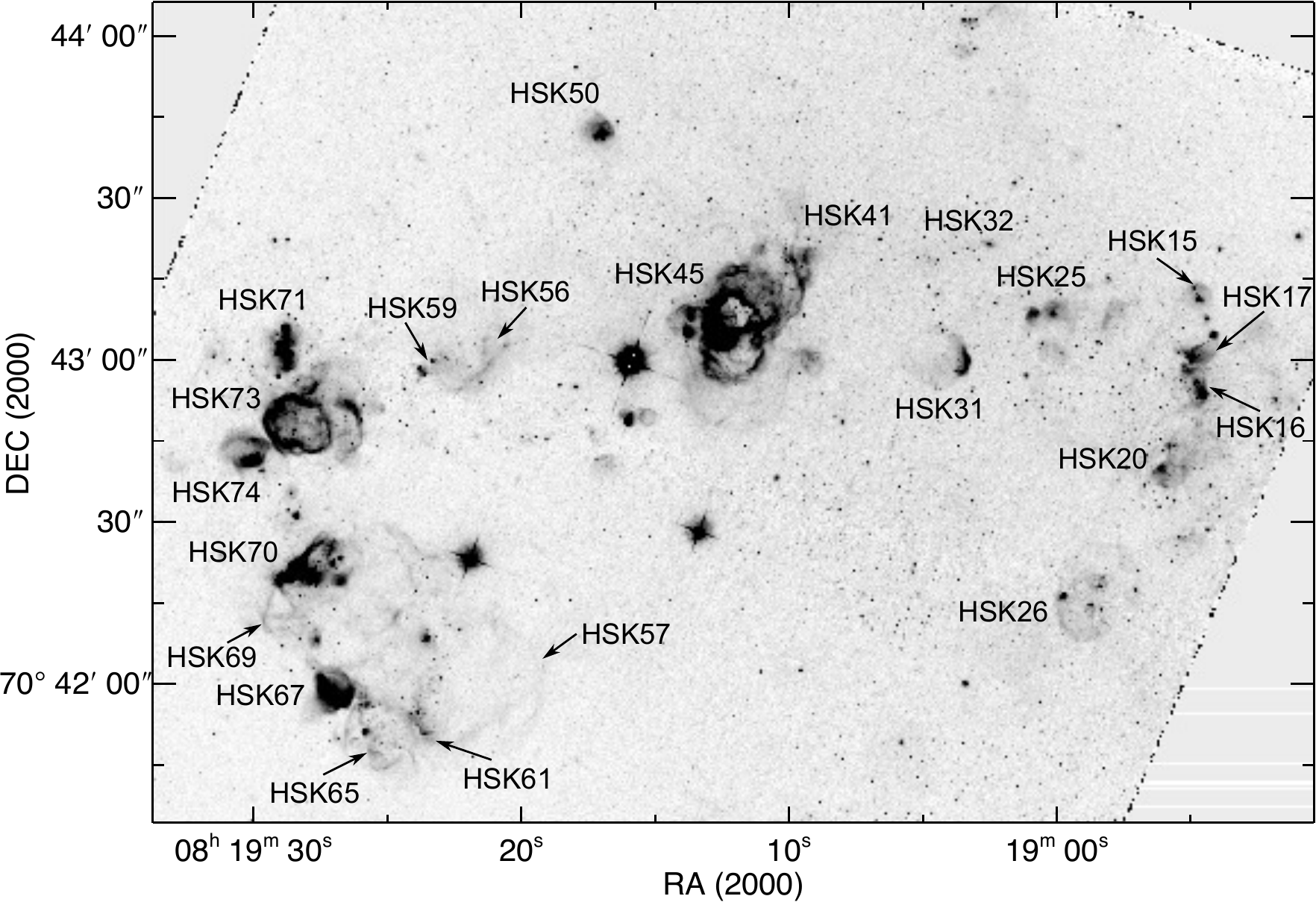}
    \caption{Top --  an \Ha image obtained by combining our deep narrow-band image and the
        sum of FPI data cube channels, with overlaid borders of the unified star formation complexes we identified (see Section~\ref{sec:complexes}) and of the internal supershell (see Section~\ref{sec:superbubble}); bottom --
        the \HST \Ha image  with HSK numbers of \HII regions denoted according to  \citet{hodge94}. A dashed line on the top panel outlines the borders of the area shown on the bottom panel.}\label{fig:reg_separation}
\end{figure}

\begin{figure}
    \includegraphics[width=\linewidth]{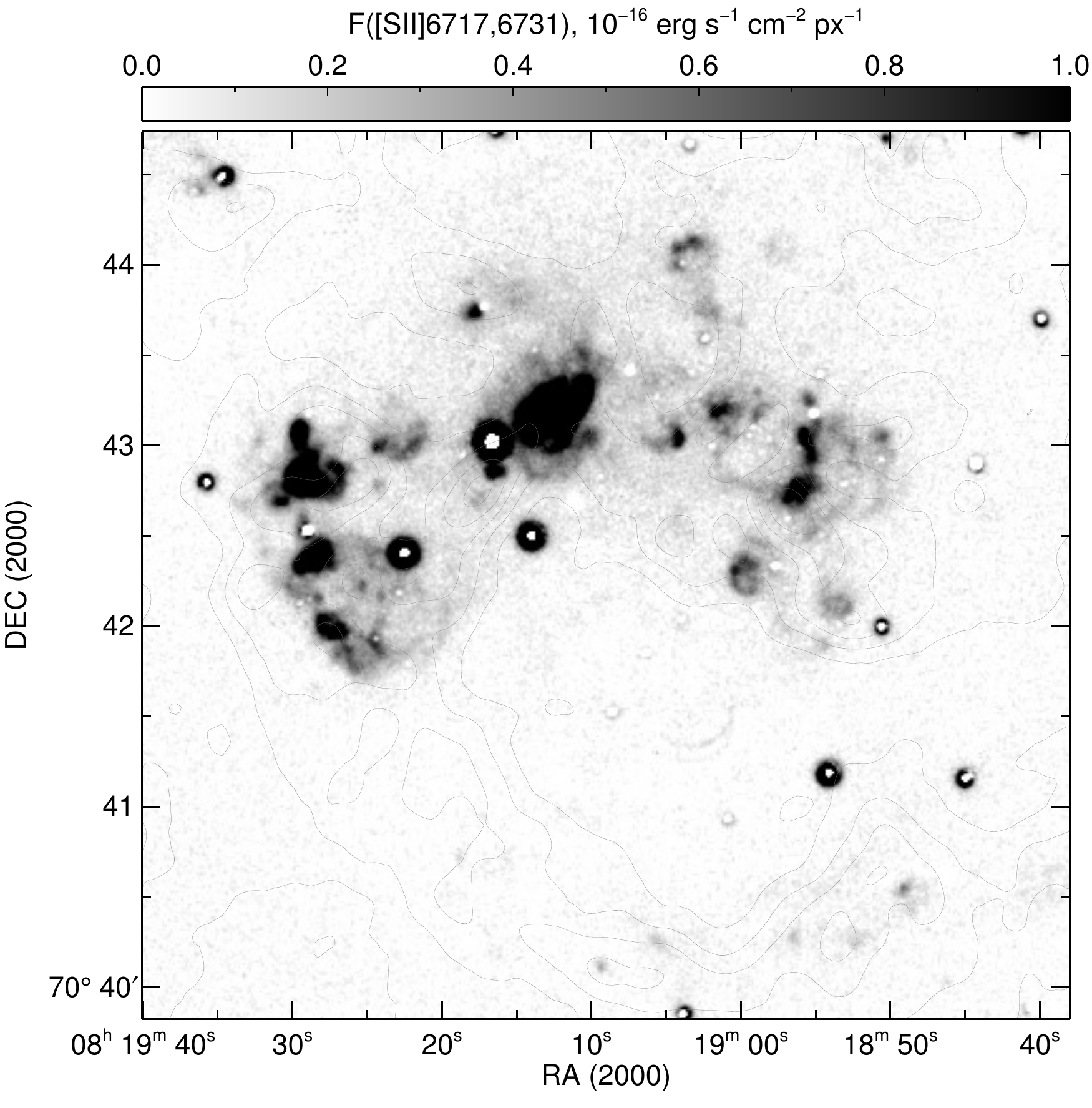}

    \includegraphics[width=\linewidth]{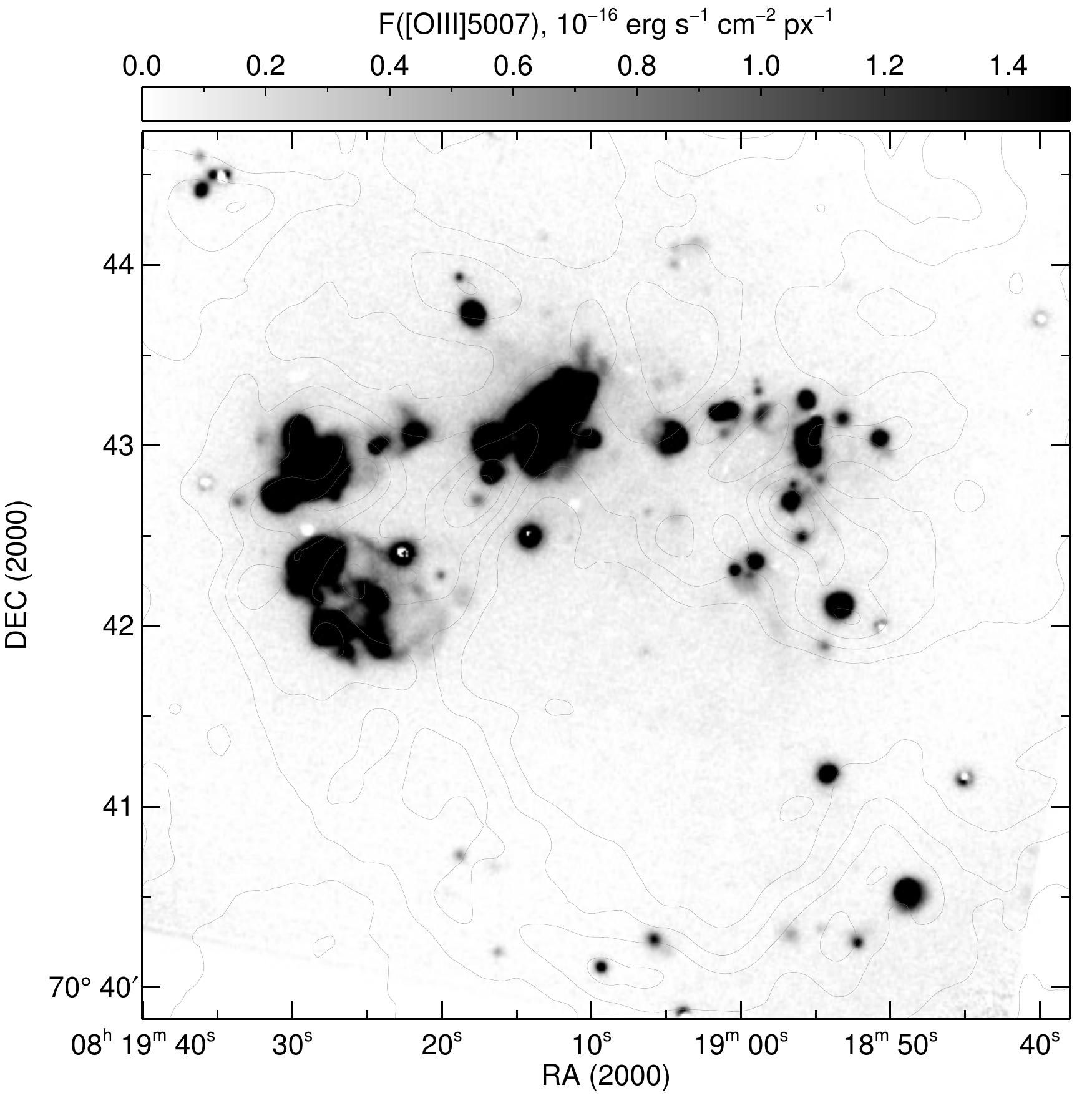}
    \caption{\SII (top) and \OIII (bottom) images obtained with narrow-band filters after continuum subtraction. Isophotes denote the \HI column density distribution.}\label{fig:s2o3img}
\end{figure}

\subsection{Narrow-band imaging}

Deep optical images in the \Ha, \SII and \OIII emission lines were obtained at the 6-m telescope with the same
SCORPIO-2 device using the filters FN655, FN674 and IFP502 with central
wavelengths of  6559 \AA, 6733 \AA\, and 5014 \AA, and FWHM = 97 \AA,  60 \AA\, and 18 \AA\, respectively.
The images taken with FN674 and IFP502 represent the
\SII 6717, 6731~\AA\, and \OIII 5007~\AA\, emission lines respectively.
As the FWHM of the FN655 filter is broader than the separation  between the \Ha and [N~{\sc ii}]\,
emission lines, the image in  this filter is contaminated  by [N~{\sc ii}] 6548, 6584~\AA\, emission.
However, given the mean \NIIHa\, ratio for \HII regions in Ho~II adopted from \cite{egorov13}, we find that
the contribution of the \NII emission to total flux is less then 4 per cent.

We used the broad-band  FN641, FN712 and SED525 filters centred on the continuum near the
\Ha, \SII and \OIII emission lines to subtract the stellar contamination from the images obtained in the same night.
Note that the subtraction was not ideal and residuals due to the stellar contribution can be seen in our
final images in several areas, especially in those taken in the \SII
lines. In order to calibrate the emission-line images to energy fluxes, we observed the
standard star AGK+81d266 (G191B2B in the case of \OIII observations) immediately after observing the galaxy.

\subsection{Archival  data used}
\label{sec_obs3}

In order to study the \HI gas kinematics in the SGS, we analysed archival \HI 21-cm line VLA data from
the LITTLE THINGS survey  \citep{littlethings}. We used the natural-weighted  data cube
and applied the same `derotation' procedure to it as for the FPI data cube.

We trace the recent star formation distribution in Ho~II by using the far UV images obtained with the GALEX
space telescope as a part of the LITTLE THINGS survey.

To analyse the distribution of ionized gas, we use the data of the \HST ACS/WFC  \Ha line observations carried out with the narrow-band filter F658N (program ID 10522, data are published in \citealt{hong13}.)  downloaded
from the \textit{Hubble} Legacy Archive\footnote{http://hla.stsci.edu/}. While our \Ha image is deeper
than \HST data, the last have better spatial resolution which is essential for understanding
the internal structure of bright \HII regions.

We adopted stellar photometry data from the ACS Nearby Galaxy Survey Treasury (ANGST) catalogue
\citep{angst} which includes the apparent magnitudes in the F606W and F814W filters for each star
in the \HST ACS/WFC field.

\section{Multi-wavelength overview of Holmberg~II}\label{sec:overview}

\begin{table}
    \caption{General parameters of Holmberg II}\label{tab:main_params}
    \begin{tabular}{lr}
        \hline
        Parameters (units)   & Value \\
        \hline
        Name$^a$ & Holmberg II; Ho II; DDO 50; \\
        & UGC 4305; PGC 23324; VII Zw 223 \\
        Centre position$^b$:        &      \\
        RA (J2000.0)       &   $08^h19^m03.7^s$  \\
        DEC (J2000.0)      &  $+70^\circ43'24$\farcs6     \\
        Morphology type$^a$        & Im   \\
        Holmberg radius R$_{26}^c$    & $267.3''$  (4.4 kpc)\\
        Distance$^c$                   & $3.39 \pm 0.20$   Mpc  \\
        Spatial scale                     &        16.4 pc arcsec$^{-1}$  \\
        M$_B$ magnitude$^c$      & $-16.71 \pm 0.16$ \\
        Colour $B-V^c$           &  $0.11 \pm 0.05$ \\
        Gas mass$^b$            &  $1.3\times10^9\ M_\odot$ \\
        Inclination                 & $49\degr^b$       \\
        & $38\degr^d$       \\
        & $27\degr^e$       \\
        Position angle$^b$                    & $176\degr$   \\
        Systemic velocity$^{b, c}$  & $157 \kms$  \\
        Rotation velocity$^b$       &  $35.7 \kms$ \\
        \hline
    \end{tabular}
    $^a$NED, NASA/IPAC Extragalactic Database \\ (http://ned.ipac.caltech.edu/)\\
    $^b$\cite{Oh11} \\
    $^c$Updated Local Volume galaxies database \citep[][http://www.sao.ru/lv/lvgdb/]{karach13}  \\
    $^d$\cite{moustakas10} \\
    $^e$\cite{HoII_incl_new} \\
\end{table}

The galaxy Holmberg II has been thoroughly studied in the whole wavelength range.
It is included in several large surveys: SINGS \citep{sings}, THINGS
\citep{things}, LITTLE THINGS \citep{littlethings}, HERACLES \citep{heracles}, KINGFISH \citep{kingfish} and Local Volume galaxies database \citep*{karach13}.
The basic parameters of the galaxy are listed in Table~\ref{tab:main_params}.

The morphology of atomic hydrogen in this galaxy is extremely
inhomogeneous. The observed large-scale comet-like asymmetry of
the \HI disc is probably caused by the ram pressure stripping due
to the hot intragroup medium \citep{bernard12}. The \HI
distribution in the galaxy disc represents a large number of
contiguous giant supershells and `holes' (most representable of them are shown in
Fig.~\ref{fig:colorIm}). Thus \citet{puche92} identified 51 giant
cavities and slowly expanding \HI supergiant shells in Ho~II,
some of which required a kinetic energy as high as  $10^{53}$ erg
for their formation. The subsequent analysis by
\citet{bagetakos11} who used more rigid criteria allowed them
to identify 39 H~{\sc i} cavities surrounded by shells in this galaxy,
the overwhelming majority of which coincide with objects from the
list by \citet{puche92}. The sizes of shells range from 0.26 to
2.11~kpc and their expansion velocities vary in the  $7 - 20
\kms$ interval, which corresponds to the kinematic age of $10 -
150$~Myr.

The  Ho~II rotation curve determined by \cite{puche92} based on \HI 21~cm line observations
shows rigid rotation in the central region with the radius of 2~arcmin which contains
the entire optical emission of the galaxy. \cite{Oh11} performed tilted-ring analysis of
the HI line-of-sight velocity field and obtained circular rotation models of the galaxy
for studying the dark matter distribution. We use  hereafter the basic kinematics parameters listed in
Table~\ref{tab:main_params} adopted from the above paper.

Note that there is still no consensus concerning the measurement
of the galaxy inclination. For instance, \citet{Oh11}
reported $i=49^\circ$;  \citet{moustakas10} found $i=38^\circ$
based on the results of surface photometry;
\citet*{HoII_incl_new} supposed an inclination of
$i\simeq27^\circ$ requiring to place the Ho~II disc above the
stability threshold.

The thickness of the gaseous disk  (about 400~pc)  varies
insignificantly with galactocentric distance \citep{banerjee11},
making  Ho~II a preferred object among other dIrr for studying
the nature of supershells. Such a large vertical scale height
allows supershells in Ho~II  to grow to large sizes before
breaking through the \HI disc.

\citet*{hodge94} identified  82 \HII regions in the galaxy;
hereafter for the sake of consistency we use  the HSK numbers from the catalog of the
above authors to denote bright  \HII regions shown in Fig.~\ref{fig:reg_separation}. The brightest
emission nebulae are mostly concentrated in the northern wall of
the most extended (about 2 kpc) \HI SGS (\#21 in the list by
\citealt{puche92} and \#17 in the list by \citealt{bagetakos11}),
which is the  subject of our study. Figure~\ref{fig:colorIm} shows that bright \HII regions form several
`chains' assembling the overall `central arc' (this term was used
by \citealt{stewart00}) in the northern wall of the SGS. The
brightest of these regions are concentrated in the eastern chain.
The brightest emission of heated dust is also observed in this
`central arc' \citep*[see, e.g.][]{hunter93, karach07, walter07,
wiebe14}. In general, the location of the  \Ha emission agrees
with the distribution of the  FUV emission in the galaxy, see
also figs.~5, 6 in \citet{stewart00} and fig.~29 in
\citet{weisz09a}. Bright regions of the FUV emission are also located
in the northern rim of \HI SGS.  Apart from them the region of weak
diffuse FUV emission extending to the southwest, three separate
FUV regions to the north of the central arc and a compact one to the
west of it are also seen in Fig.~\ref{fig:colorIm}. The last one
(furthest FUV region in Ho~II) was analysed by \citet*{hunter16}.

\citet{stewart00} studied  45 \HII regions using \textit{B}, \Ha
and FUV photometry and estimated the ages of the corresponding
star-forming regions by dividing them into four age groups. Most
of the star-forming regions fell into groups with ages smaller
than  6.3 Myr. This is consistent with the  star formation
history of Ho~II: according to \citep{weisz08}, the star-formation
rate has been increasing over the last several million years.

Radio observations of the galaxy  \citep*{tong95, braun07, heald09} revealed continuous
radio emission in the region of bright emission nebulae. The above authors associated this radio emission
with active star formation and suspected the presence of supernova remnants
in the eastern chain of  \HII regions. We discuss this subject in Section~\ref{sec:SNe} in more detail.
According to the \textit{Chandra} observatory data, the ultraluminous X-ray source (ULX) Ho~II X-1 located
in the eastern chain of star-forming regions is  the only X-ray source in the galaxy and
one of the sources of the observed bright synchrotron emission.

Earlier, the kinematics of ionized gas in  Ho~II was studied in
detail in the neighborhood of  ULX Ho II X-1 only
(\citealt{lehmann05}, see also our paper \citealt*{egorov16} about this object). \citet{tomita98} analysed the kinematics of
other bright  \HII regions of the galaxy based on observations
made with a slit spectrograph. They combined seven
obtained spectra into a broad strip covering only one third of
the bright  \HII regions in the SGS. The resulting
`position-velocity' diagrams (PV diagrams) for the  \HII region
that the authors denoted as the `centre' (HSK~45) revealed the
radial-velocity gradient amounting to  $30 \kms$ over 200~pc; the
most blue-shifted \Ha velocity is smaller than the \HI velocity
by about $15 \kms$.

Observations of Holmberg~II with FPI were performed previously at the Observatoire du mont
M\'egantic 1.6-m telescope as a part of \Ha kinematics follow-up survey of the SINGS sample \citep{dicaire08}.
The aim of that project was to investigate the global kinematics of galaxies observed. The
authors did not provide the kinematic parameters of Holmberg~II due to its weak
rotation and a lack of spatial coverage.

\section{Search for expanding shells in star-forming regions}\label{sec:shells}

\begin{figure*}
	\includegraphics[width=\linewidth]{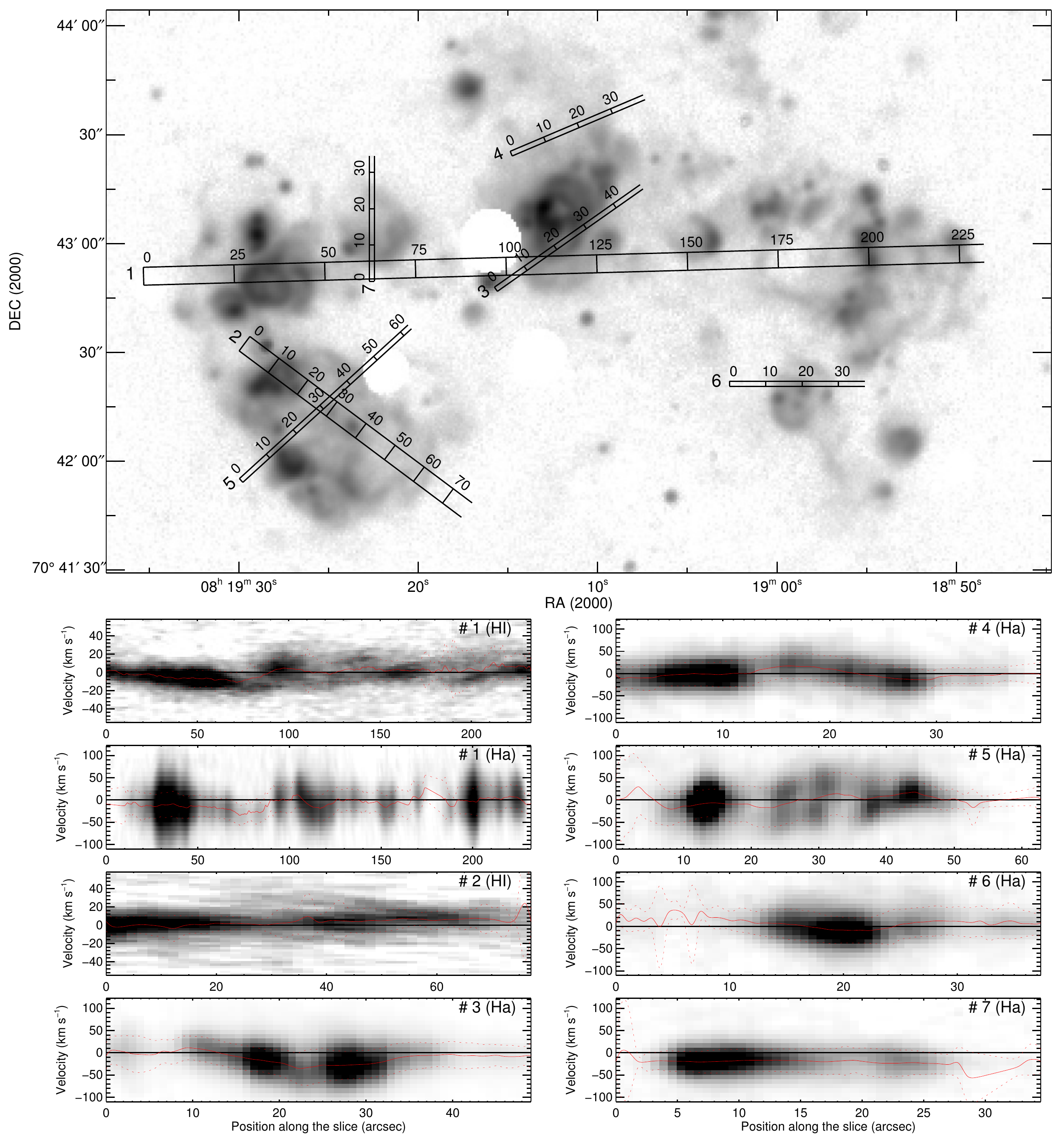}
	\caption{Examples of H$\alpha$ and HI PV-diagrams. Top panel: an \Ha image
		arcsinh (intensity-scale) with the localization of the PV diagram superimposed.
		PV diagram names shown to the left of the bar denote their position. Bottom panels: examples of
		PV diagrams with the median velocity (red solid curves) and its dispersion (red dotted curves) overlaid.
		The zero-velocity line is shown by a black solid line.}\label{fig:pv}
\end{figure*}

\begin{figure}
	\includegraphics[width=\linewidth]{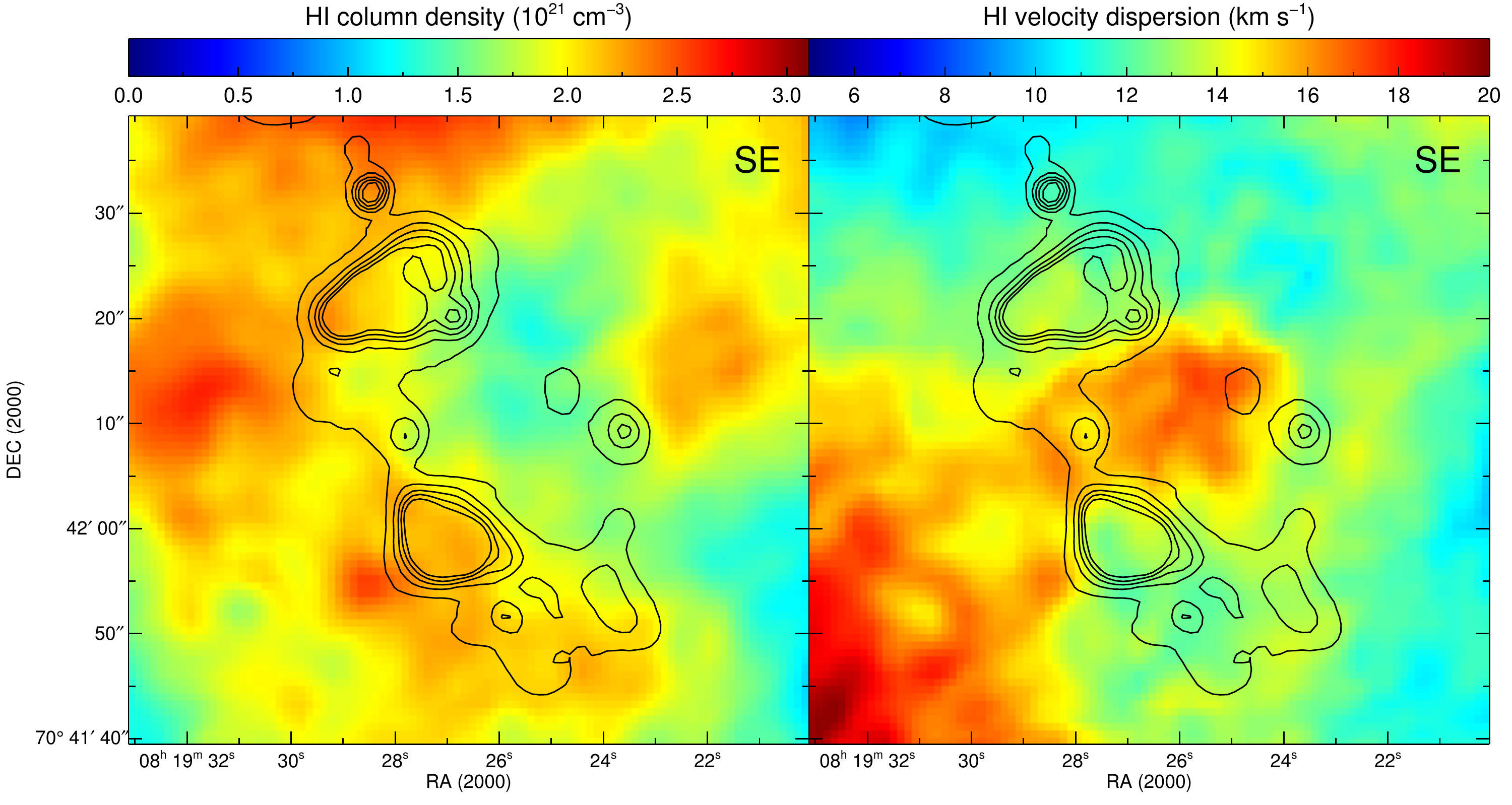}
	
	\includegraphics[width=\linewidth]{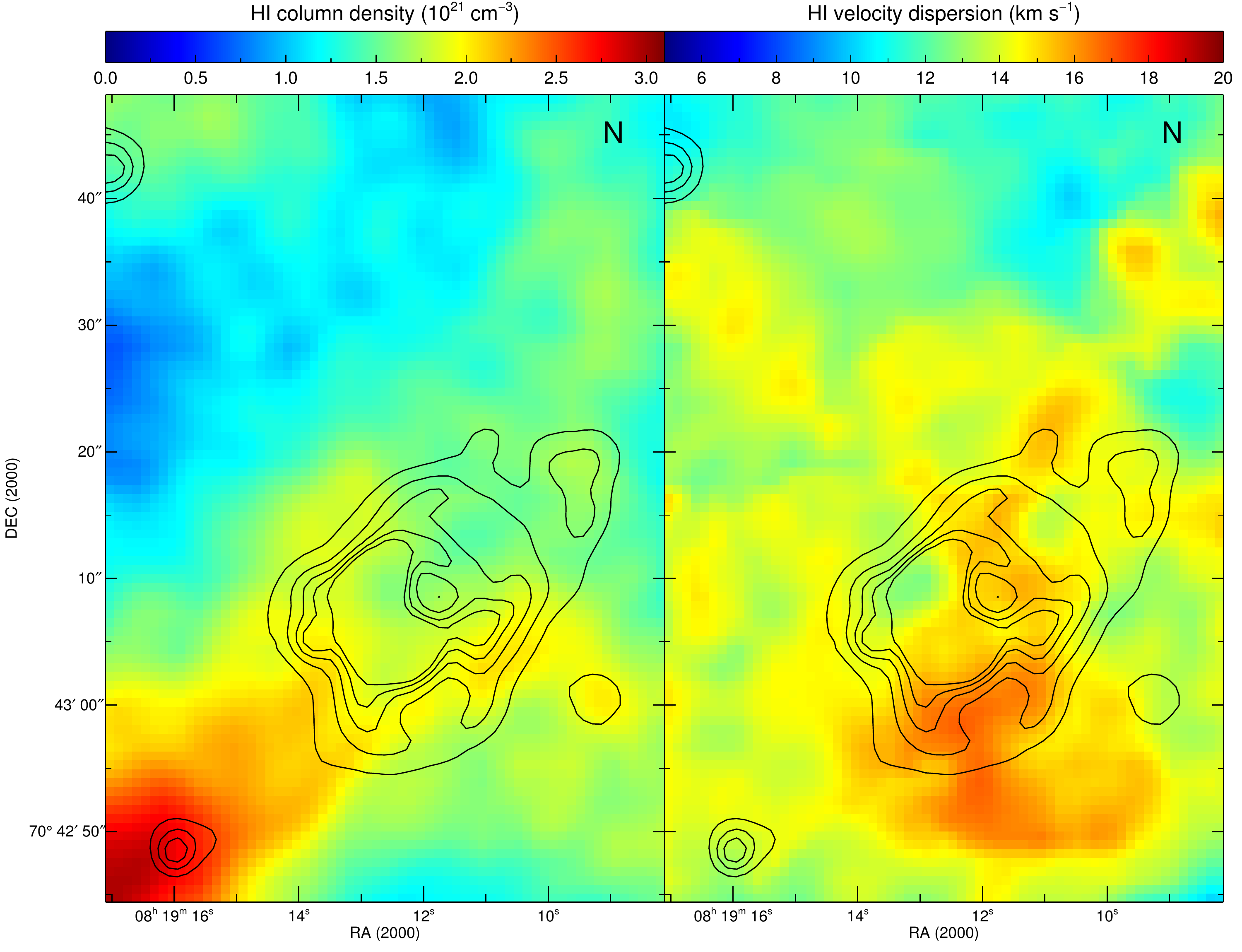}
	
	\includegraphics[width=\linewidth]{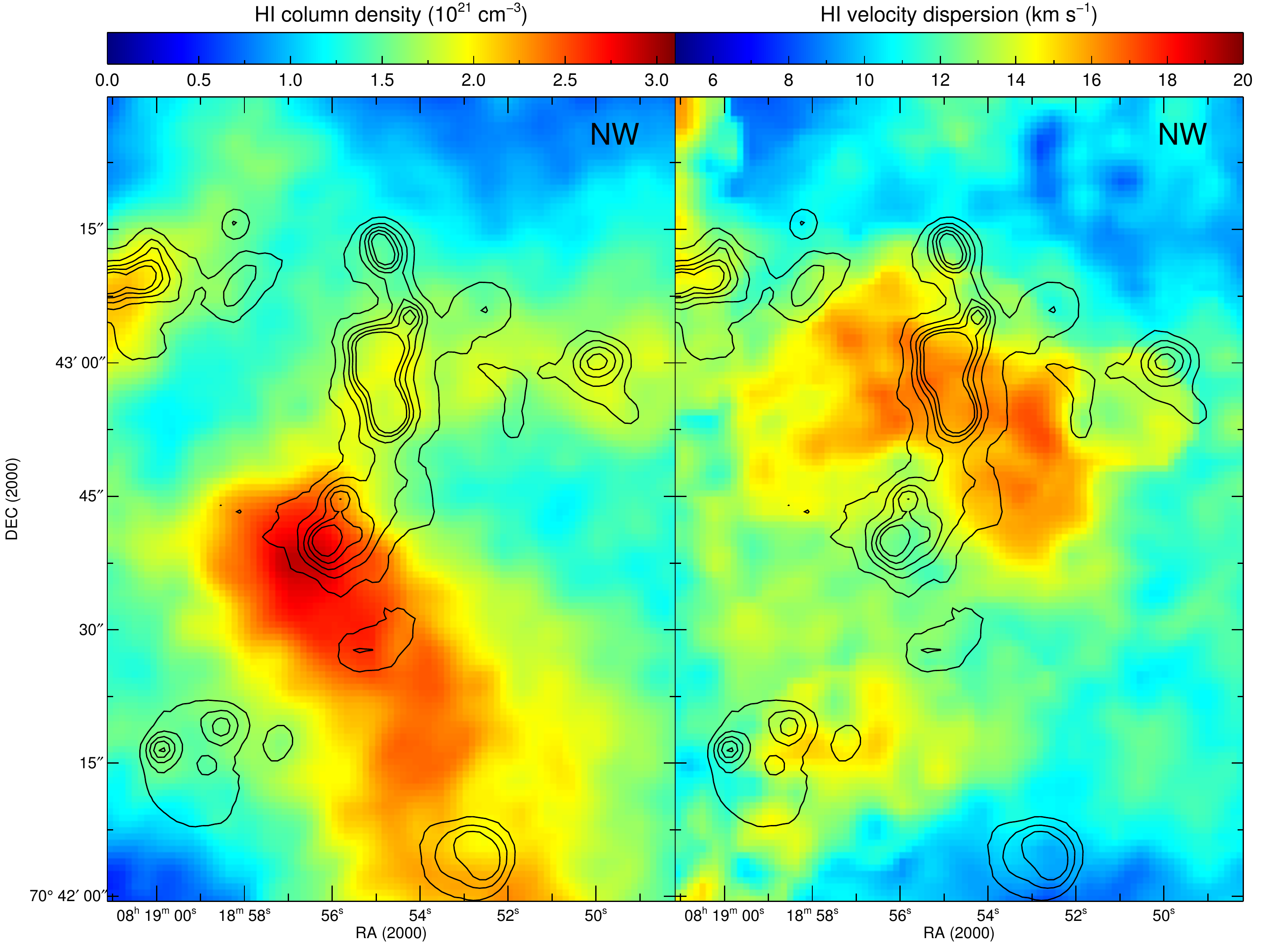}

	\caption{\HI column density (left) and line-of-sight velocity dispersion (right) maps of three
		star-forming complexes (SE, N and NW) with local \HI superbubbles detected. \Ha line intensity
		contours are overlaid.}\label{fig:HI_maps}
\end{figure}

\begin{figure}
	\includegraphics[width=\linewidth]{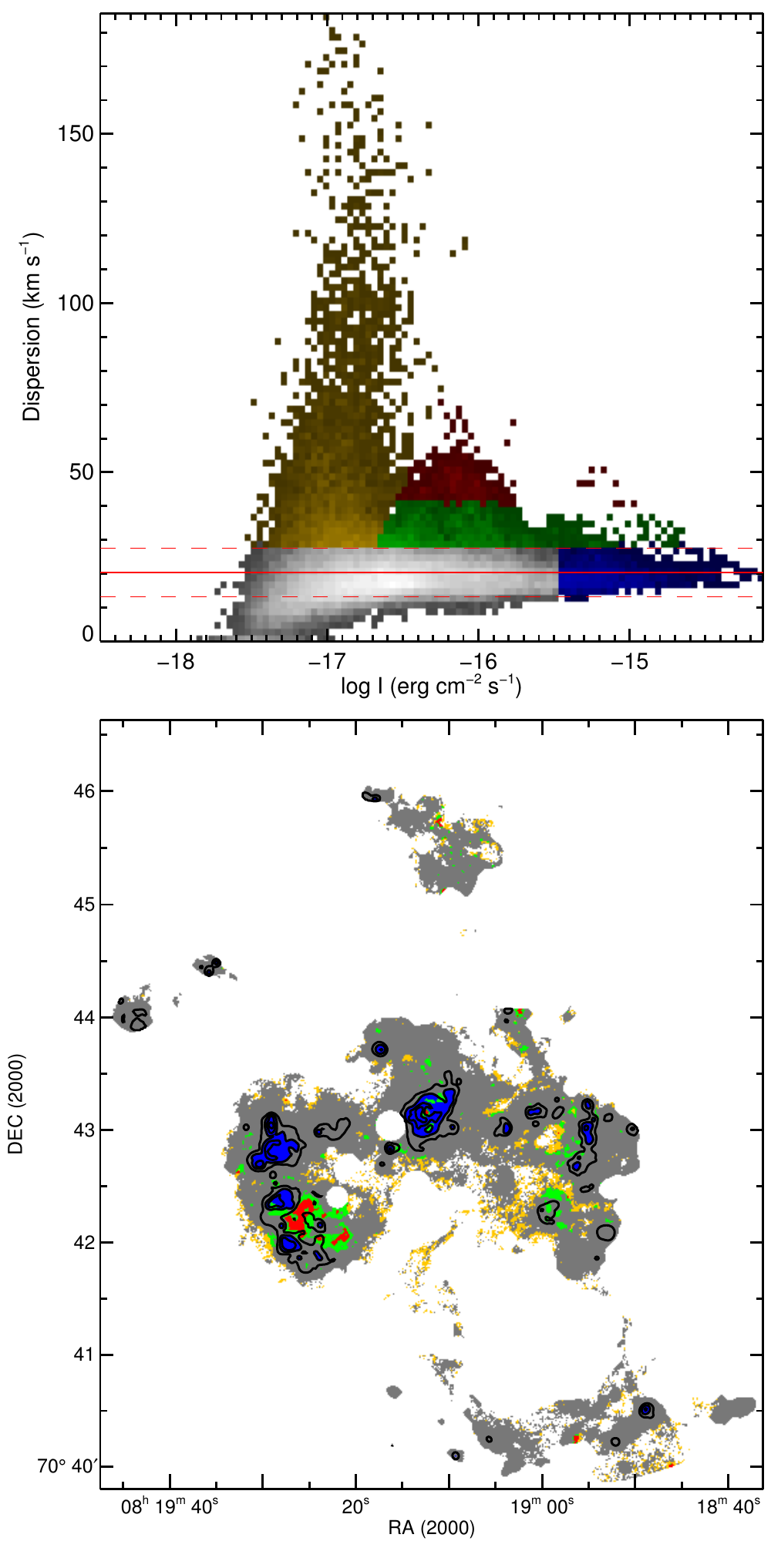}
	\caption{\Ha intensity -- velocity dispersion diagram for Holmberg II (top). A red solid horizontal line marks the intensity-weighted mean velocity dispersion of ionized gas in Ho~II and the dashed
		lines -- its standard deviation. The bottom plot shows the location of regions marked by different colours in the $I - \sigma$ diagram. }\label{fig:i_sigma}
\end{figure}

Our deep \Ha images of Holmberg~II and the archival {\it HST} images of the galaxy
(Fig.~\ref{fig:reg_separation}) reveal numerous faint extended filaments connecting
several \HII regions and located beyond the bright nebulae. A similar picture is observed in the \SII and \OIII emission lines (see Fig.~\ref{fig:s2o3img}).
Based on this morphology, we combined the emission features observed in the rim of the SGS
into four extended areas (SE, NE, N and NW in Fig.~\ref{fig:reg_separation}) and into two areas
beyond the SGS (ExtN and ExtNE).

To confirm the conclusion that the observed emission structures we identified are indeed unified physically-bound complexes of ongoing star formation, we performed a detailed
analysis of the kinematics of ionized and neutral gas in these areas.

The structure and kinematics of both ionized and neutral gas in the studied \HI SGS are not homogeneous.
Figure~\ref{fig:pv} shows several  examples of PV
diagrams in the \Ha and \HI lines, that demonstrate how line-of-sight velocity changes with position along
a chosen direction. Deviations of the line-of-sight velocity from the circular rotation model of the galaxy
(zero velocity line) are clearly seen in the largest slice \#1. Signs of the `velocity ellipses' --
the characteristic feature of expanding shells -- are also present in this PV diagram.

\begin{table}
	\caption{Expansion velocity, size and kinematic age of faint shell-like structures of ionized gas shown in Figs.~\ref{fig:SE_shells}--\ref{fig:sb_stars}. Ages of the bright \HII regions tied with these shells are also shown.}\label{tab:shells_summary}
	\begin{tabular}{cccc}
		\hline
		Shell & $V_{exp}, \kms$ & Size, pc & Age, Myr \\
		\hline
		SE1 & 15 & 86 & 1.7\\
		SE2 & 30--37 & $322\times230$ & 2.6\\
		SE3 & 25--30 & $282\times188$ & 2.8\\
		SE4 & 27--32 & $190\times128$ & 1.8\\
		SE5 & 27 & $358\times240$ & 4.0\\
		SE6 & $\simlt20$ & 74 & 1.1\\
		\multicolumn{3}{c}{Age of HSK~67, 70:} &  $3.8 - 4.8^*$ \\
		\multicolumn{3}{c}{} &  $2.5 - 4.5^{**}$ \\
		\hline
		NE1 & 26 & $72\times50$ & 0.8\\
		NE2 & 15 & $162\times104$ & 3.2\\
		NE3 & 17 & $254\times224$ & 4.5\\
		NE4 & 24 & $210\times168$ & 2.6\\
		\multicolumn{3}{c}{Age of HSK~59, 71, 73:} & $3.5 - 4.1^*$ \\
		\multicolumn{3}{c}{} &  $2.5 - 4.5^{**}$ \\
		\hline
		N1 & 22 & $362\times316$ & 4.9\\
		N2 & 28 & $142\times96$ & 1.5\\
		N3 & 10 & $124\times76$ & 3.7\\
		N4 & 18 & $230\times203$  & 7.2\\
		\multicolumn{3}{c}{Age of HSK~45:} &  $3.7^*$  \\
		\multicolumn{3}{c}{} &  $2.5 - 3.5^{**}$ \\
		\hline
		NW1 & 25 & $392\times312$ & 4.7\\
		NW2 & 21 & $112\times70$ & 1.6\\
		NW3 & 21 & $163\times102$ & 2.3\\
		NW4 & 27 & 216 & 2.4\\
		NW5 & $\simeq 20 $ & 226 & 3.4\\                
		\multicolumn{3}{c}{Age of HSK~15, 16, 17, 20, 25:} & $4.8 - 6.2^*$ \\
		\multicolumn{3}{c}{} &  $4.5 - 6.3^{**}$ \\
		\hline
		ExtNE1 & 20 & $49$ & 1.5\\
		\multicolumn{3}{c}{Age of ExtNE:} &  $3.5 - 4.5^{**}$ \\
		\hline
		ExtN1 & 23 & 139  & 3.6 \\
		\multicolumn{3}{c}{Age of ExtN:} &  $3.5 - 4.5^{**}$ \\
		\hline
		S1 & 40  & 120  & 1.8\\
		\hline
		\multicolumn{4}{l}{\begin{scriptsize}
				$^*$Estimates made by \citet{wiebe14} using H$\beta$ equivalent width.\end{scriptsize}}\\
		\multicolumn{4}{l}{\begin{scriptsize}
				$^{**}$Estimated by \cite{stewart00} using \Ha and FUV data.
			\end{scriptsize}} \\
			
		\end{tabular}
	\end{table}

\begin{table*}
	\caption{Sizes and ionization budget of complexes of ongoing star formation}\label{tab:ionization}
	\begin{tabular}{ccccccc}
		\hline
		Complex & Size & F(\Ha),  & N(OV5) & N(OB)  & $\mathrm{Q_{H\alpha}}$ & $\mathrm{Q_{stars}}$
		\\
		& pc & $10^{-13}$ erg s$^{-1}$ cm$^{-2}$ & & &  $\mathrm{10^{50}\ s^{-1} }$&$\mathrm{10^{50}\ s^{-1} }$\\
		\hline
		SE &  $1150\times1050$ &9.3 & 56 & 35 & 8.8 &9.7\\
		NE & $1400\times980$& 9.7 & 58 & 26 & 9.1 &8.3 \\
		N &  $1500\times1000$ &14.0 & 83 & 36 & 13.2 & 11.1\\
		NW & $2450\times1600$ & 8.4 & 50 & 31 & 7.9 &7.6 \\
		Int.shell (diff) & $2730\times2280$ &  0.7 & 4 & 5 & 0.7 &1.0\\
		Int.shell (all) & $2730\times2280$ & 1.7 & 10 & 7& 1.6 & 1.5\\
		ExtN & $1530\times1030$ & 1.1 & 7 & $8^*$ & 1.1 & $1.9^*$\\
		ExtNE &  $960\times700$ & 0.5 & 3 & 3 & 0.5 &1.1\\
		\hline
		\multicolumn{6}{l}{\begin{scriptsize}
				$^*$Might be underestimated because of the incomplete coverage by \HST observations (see Fig.~\ref{fig:colorIm}).\end{scriptsize}}\\
		
	\end{tabular}
\end{table*}

Earlier, \cite{weisz09a} identified four local \HI shells in the SGS wall with sizes of $300 - 400$~pc
 expanding with velocities  $7-15 \kms$. These shells (\# 8, 12, 14 and 27 in Fig.~2 of their paper)
are located in the immediate vicinity of the \HII HSK~70, 16, 17,
and 26 regions (see fig.~\ref{fig:reg_separation}). In addition,
shells \# 16 and 23 are also seen and located north of the
central chain of star-forming regions which might be connected
with it.

We refined the location of the \HI shells associated with 
star-forming regions and performed search for ionized
gas shells in the galaxy. Below, we report the details and results
of our kinematic analysis.

An expanding shell can be identified by a velocity ellipse (or a part of it) in PV diagrams or by 
 splitting of an emission-line profile. In the case where the expansion velocity of a shell is small
and  spectral resolution is insufficient to resolve two
components, the shell distinguishes itself by its higher velocity
dispersion in the centre.

We used all the above techniques to search for expanding shells and estimate their expansion velocities.

Figure~\ref{fig:HI_maps}  shows the \HI column density and line-of-sight velocity dispersion
(second moment) maps for  three areas -- SE, N and NW -- where we see increased \HI velocity dispersion.
The areas of high velocity dispersion in Fig.~\ref{fig:HI_maps} correspond indeed to the expanding \HI superbubbles.
This suggestion is confirmed by our analysis of PV diagrams. For example, the `velocity ellipse' might be
seen at positions of $100 - 140$ , $170 - 220$ arcsec along PV diagram \# 1 (HI) and $15 - 65$ arcsec along
PV diagram \# 2 (HI) in Fig.~\ref{fig:pv}, which correspond to the regions of the increased velocity dispersion in
Fig.~\ref{fig:HI_maps} in the areas of N, NW and SE respectively.

It was shown in many studies \citep[][see also references
therein]{MunozTunon1996,ML12} that the intensity -- velocity
dispersion diagrams of ionized gas ($I - \sigma$) can be successfully used
to identify areas, where the increased velocity dispersion
is caused by expanding shells. We fitted the  \Ha line profile to a
single-component Voigt profile in each pixel obtained from the
FPI data cube and constructed the  $I - \sigma$ diagram for the galaxy Holmberg~II (see Fig.~\ref{fig:i_sigma}). The red line in
the diagram marks the intensity-weighted average velocity
dispersion of ionized gas in the galaxy -- $\sigma_m = 20.4 \pm
7.1 \kms $. Identical colours in the diagram (the top panel) and
map (the bottom panel) are used to highlight the characteristic
areas. Thus, the horizontal strip with relatively low velocity
dispersion and high surface brightness  marked in blue
corresponds to bright  \HII regions and includes  50 per cent of
the galaxy's \Ha flux. Areas of increased dispersion, which are
most likely associated with spectroscopically unresolved
expanding shells, are shown in green colour, whereas
the red colour corresponds to shell-like structures with clear
separation of the components of the \Ha line profile. The remaining
areas of the increased velocity dispersion are shown in orange
colour, and areas of low surface brightness, in grey colour.

We performed a more bona fide search for expanding structures of ionized gas by analysing the PV diagrams that uniformly cover each region of ongoing star formation in the galaxy.
Some examples of them are shown in Fig.~\ref{fig:pv}. In the areas shown in red colour in Fig.~\ref{fig:i_sigma},
we see clear evidence for gas expansion in the form of the velocity ellipse or a part of it.
Subsequent fitting of the  \Ha profiles by one, two or three Voigt components
allowed us to estimate the expansion velocities of the corresponding structures.

Our analysis of gas kinematics in the \HI and \Ha lines revealed
three local neutral-gas shells associated with star-forming regions and 22 ionized gas shells. Their
location is shown in grey and blue colours respectively in Fig.~\ref{fig:SE_shells} -- \ref{fig:sb_stars}.

We found no signs of  expansion of bright  \HII regions.
Despite  a rather high spectral resolution of  FPI, the
expansion  velocities of these regions proved to be low: it does
not exceed $11 \kms$ (half of the instrumental FWHM of FPI used).
All the expanding shells of ionized gas, which we found,
 represent faint filamentary features in the \Ha image and
 possibly reside in a medium with a relatively low
density. In our earlier paper \cite{wiebe14}, we have estimated the
expansion velocity of the \HII regions HSK45 and HSK73, however,
it is now clear that the evidence of their expansion does not
refer to bright nebulae but rather to faint shell-like
ionized-gas structures around them.

We estimated the kinematic ages of the identified local \HII shells in terms  of \citet{weaver77}
model by the following relation:
\[
t=0.6 R/V_{\rm exp},
\]
where $R$ is the radius in pc,  $V_{\rm exp}$ -- the expansion velocity in $\kms$ and $t$ -- 
the age in Myr.

To estimate the ages of bright  \HII regions that show no signs of its expansion,  in \citet{wiebe14}  we used the equivalent  width (EW) of the  $H_{\beta}$ emission line which correlates rather well with the age of
\HII regions \citep[see, e.g.][]{schaerer98, leitherer99}.

The derived ages of the faint local shells found inside the \HII complexes and of bright \HII regions are summarized in Table~\ref{tab:shells_summary}.  This table also lists the age estimates made by \citet{stewart00} for star-forming regions based on the results of FUV and optical observations.  Because of the uncertainty in the method used for age determination, the above authors
divided all \HII regions
studied into four age groups:  $0-3.5$ Myr, $3.4-4.5$ Myr, $4.5-6.3$ Myr and older than 6.3 Myr
(see table~5 in their paper).

    Given the results of the search for expanding shells and the study of their kinematics, we can conclude
    that the observed emission structures in the four areas in the SGS walls (SE, NE, N and NW) that
    we identified based on morphology and in two areas outside the SGS (ExtN and ExtNE) are indeed
    physically connected. This means  (see also Sections~\ref{sec:morphology} and \ref{sec:complexes} below)
    that the current star formation in the galaxy is represented by unified complexes with sizes of several
    hundred pc which combine several bright nebulae.

\section{Morphology and ionization budget of star formation complexes}\label{sec:morphology}

As shown above, the deep  \Ha images of the galaxy and the presence of local \HI and \HII shells
suggest the existence of unified extended complexes joining several bright HSK nebulae. Hence
the four extended areas (SE, NE, N and NW in Fig.~\ref{fig:reg_separation}) that we identified in the SGS studied and the  areas outside this SGS (ExtN and ExtNE) are physically connected
complexes of ongoing star formation.

Each of the star forming complexes contains several young star clusters.  \cite{weisz09a}
and \cite{bastian11} compared observed colour-magnitude diagrams based on \HST data to
stellar evolution tracks and selected objects younger than 10 Myr that should be OB stars. We used the
same method as \cite{weisz09a} to obtain the list of OB stars with their magnitudes from the
Hubble source catalogue (HSC\footnote{http://archive.stsci.edu/hst/hsc/}). The locations
of the OB stars identified in such a way agree well with the list reported by \cite{bastian11}.
We used F555W- and F814W-band photometry to estimate the $V$-band absolute magnitude of each
star. We further assume that all the selected stars are young main-sequence stars (the method used to select
them should ensure this) and use O-type star models from \cite*{Martins05} to compute the bolometric
luminosity and the number of ionizing photons $Q_0$ from each star. These stars are shown in
Figs.~\ref{fig:SE_shells}--\ref{fig:sb_stars} by asterisks with the size proportional to their
luminosity. We used the above procedure to estimate the amount of ionizing quanta $Q_{stars}$ in Table~\ref{tab:ionization} from the OB stars observed in each complex.

Similar values can be readily inferred from the observed number of H$\alpha$ photons.
With the observed \Ha flux of each complex (corrected for interstellar extinction $A_V=0.11^m$), we estimated the equivalent number of O5V stars (N(O5V) in Table~\ref{tab:ionization}) and of ionizing photons $Q_0$ needed for gas ionization in the area with the procedure described in \cite{osterbrock}. For the case of optical thick nebulae, its \Ha luminosity $L(\mathrm{H}\alpha)$ depends on $Q_0$ as:
\begin{equation}
\frac{L(\mathrm{H}\alpha)}{h\nu_{\mathrm{H}\alpha}} \simeq \frac{\alpha^{eff}_{\mathrm{H}\alpha}}{\alpha_B}Q_0 \simeq 0.45Q_0,
\label{eq:ionization}
\end{equation}
where the \Ha effective recombination coefficient $\alpha^{eff}_{\mathrm{H}\alpha} \simeq 1.17\times10^{-13}\ \mathrm{cm^{3}\ s^{-1}}$ and the total recombination coefficient of hydrogen $\alpha_B \simeq 2.59\times10^{-13}\ \mathrm{cm^{3}\ s^{-1}}$ for $T = 10\,000$~K. Further, we will refer the estimated amount of needed ionization photons $Q_0$ from (\ref{eq:ionization}) as $Q_\mathrm{H\alpha}$ to mention that it has been converted from the \Ha luminosity.

In such a way, our measurements of the \Ha flux of \HII regions and of the luminosity of OB stars inside them were used to estimate both the needed and available amount of ionizing quanta in all star formation complexes studied.  The results are summarized in Table~\ref{tab:ionization}.

Let us now consider the structure, kinematics and energy sources in each selected  \HII complex in
more detail.

\subsection*{SE complex (HSK 57, 61, 65, 67, 69 and 70)}

\begin{figure*}
    \includegraphics[width=\linewidth]{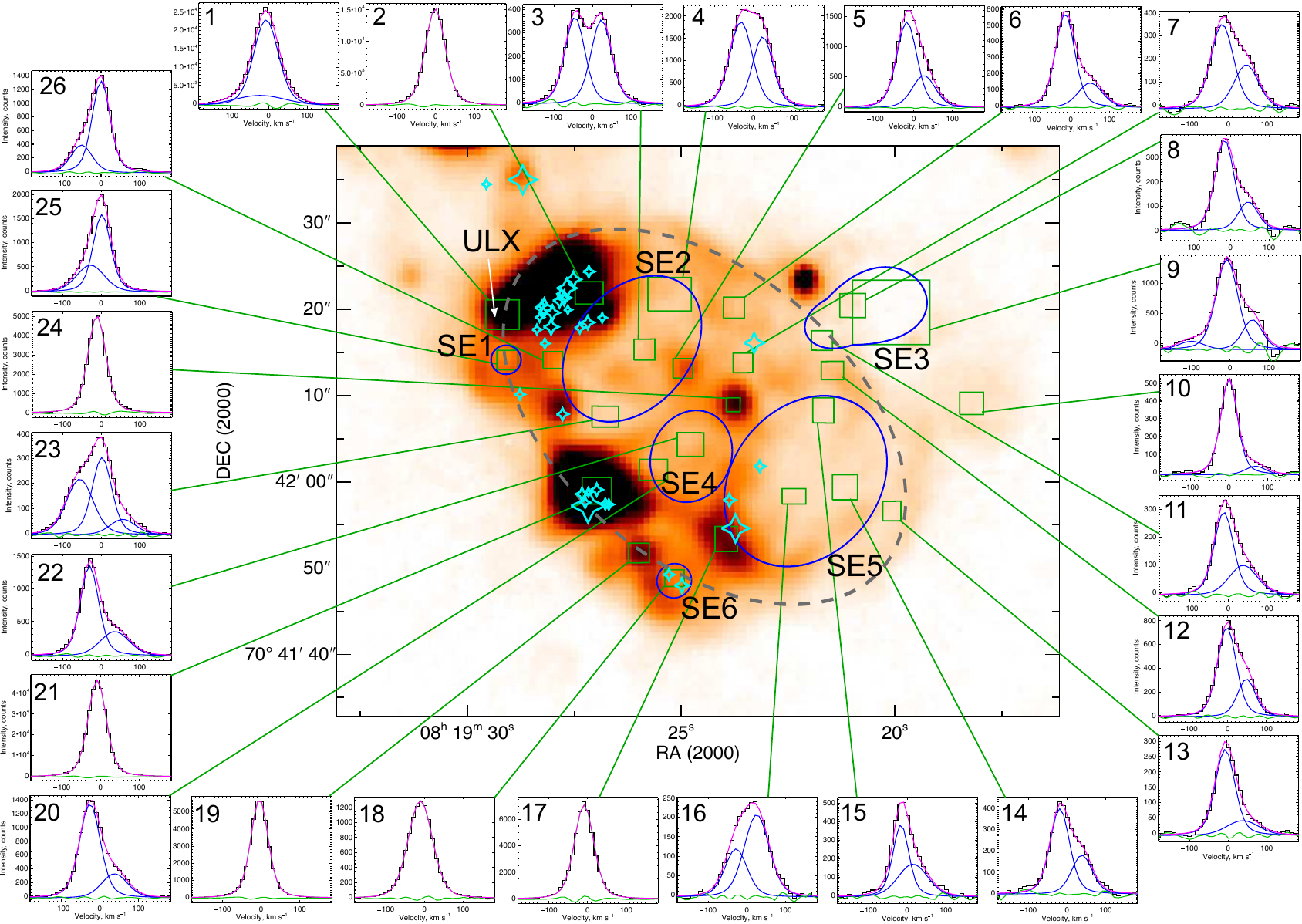}
    \caption{\Ha image of the SE complex; \Ha line profile examples obtained by integration
        inside the green borders in the image. The results of 1, 2  or 3 component Voigt fitting are shown:
        blue colour denotes the individual components, while the fitting model is shown in magenta colour.
        Green colour is used to show the residuals after  subtraction of the model from the
        observed line profile.
        Blue ellipses in the image mark the positions of faint shell-like structures identified
        in the area. A grey dashed ellipse shows the location of the neutral-gas shell identified as a result
        of the analysis of the \HI PV diagrams and associated with the observed multishell complex of ionized
        gas. Light blue asterisks denote the location of OB-stars identified by the analysis of \HST colour--magnitude diagrams; their sizes are proportional to their luminosity. }\label{fig:SE_shells}
\end{figure*}

\begin{figure*}
	\includegraphics[width=\linewidth]{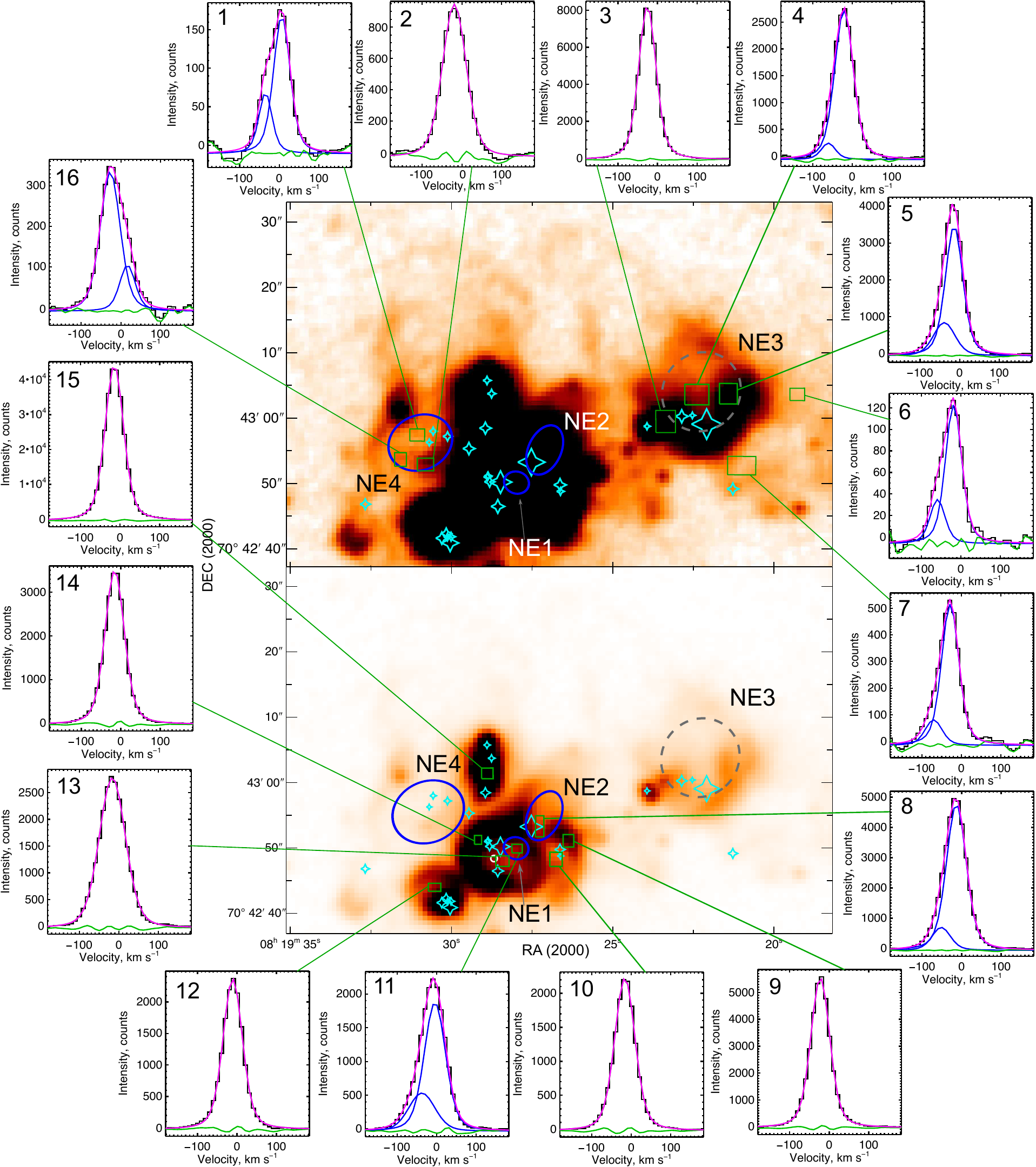}
	\caption{The same as in Fig.~\ref{fig:SE_shells}, but for the NE complex.  The top and bottom images
		correspond to the same area with different contrast. White circle shows the area of \HeII 4686~\AA\, emission detected in \citet{egorov13}.}\label{fig:NE_shells}
\end{figure*}

\begin{figure*}
	\includegraphics[width=\linewidth]{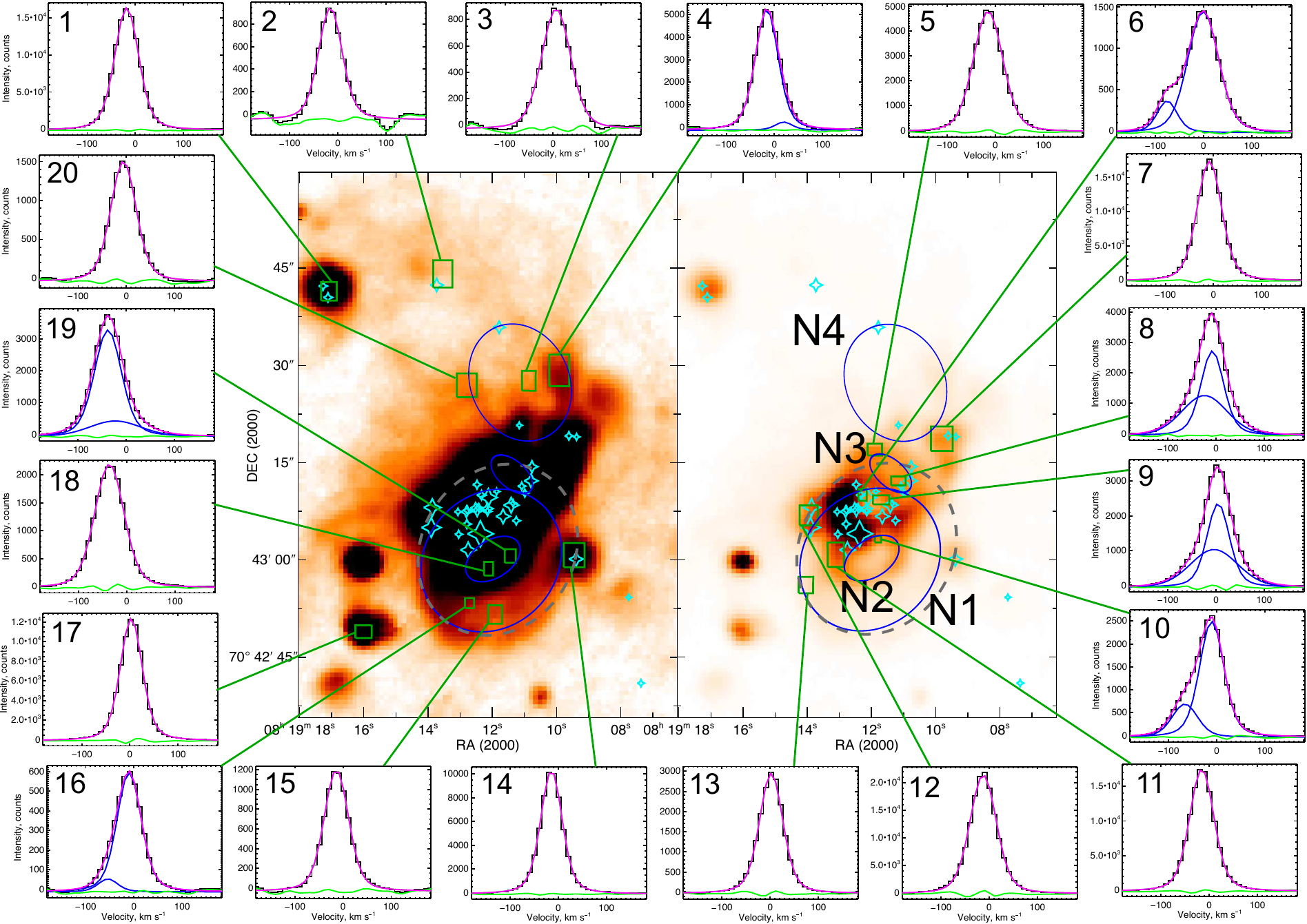}
	\caption{The same as in Fig.~\ref{fig:SE_shells} but for the N complex.  Right and left images
		correspond to the same area with different contrast.}\label{fig:N_shells}
\end{figure*}

\begin{figure*}
	\includegraphics[width=\linewidth]{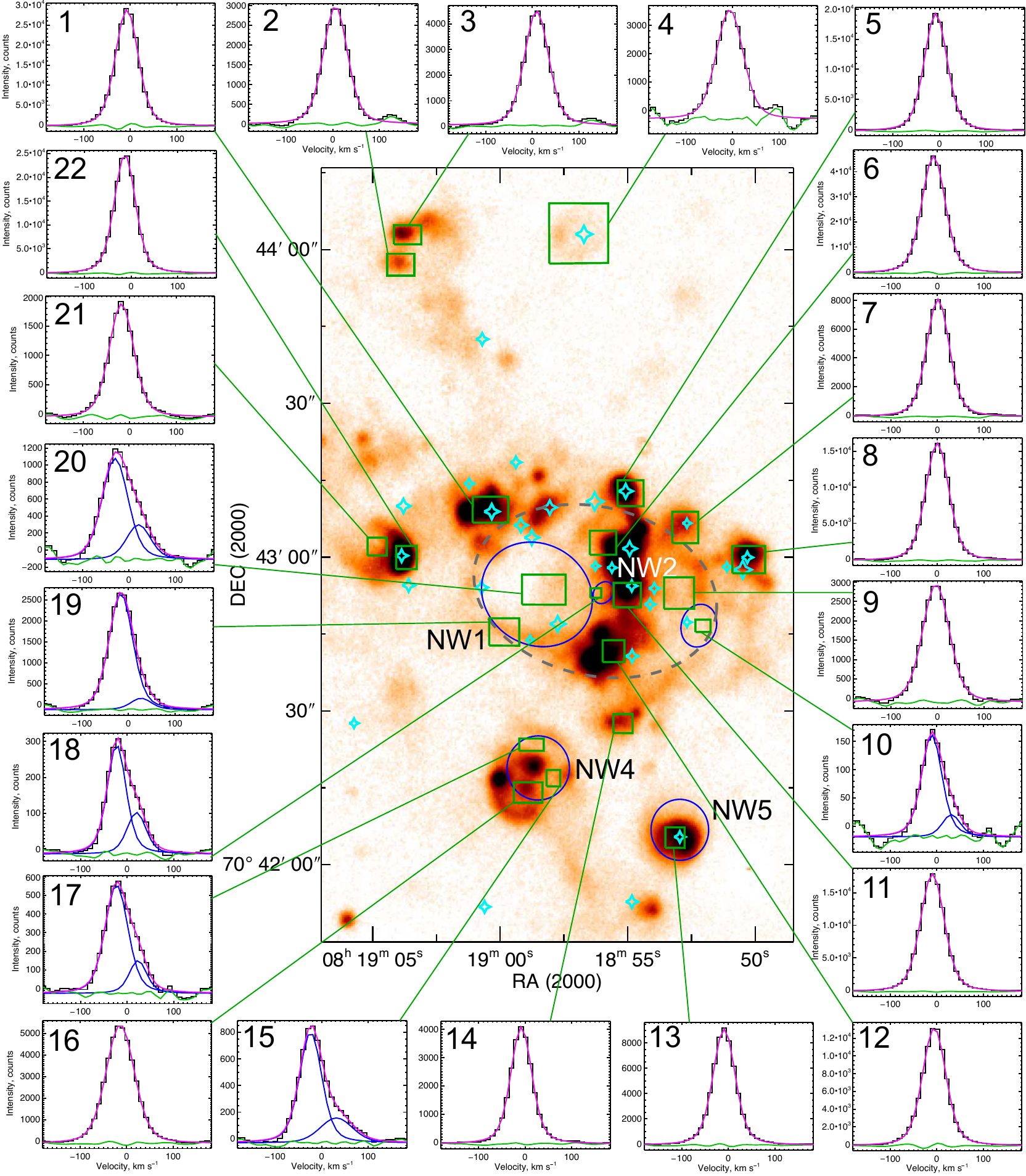}
	\caption{The same as in Fig.~\ref{fig:SE_shells} but for the NW complex.}\label{fig:NW_shells}
\end{figure*}

\begin{figure*}
	\includegraphics[width=\linewidth]{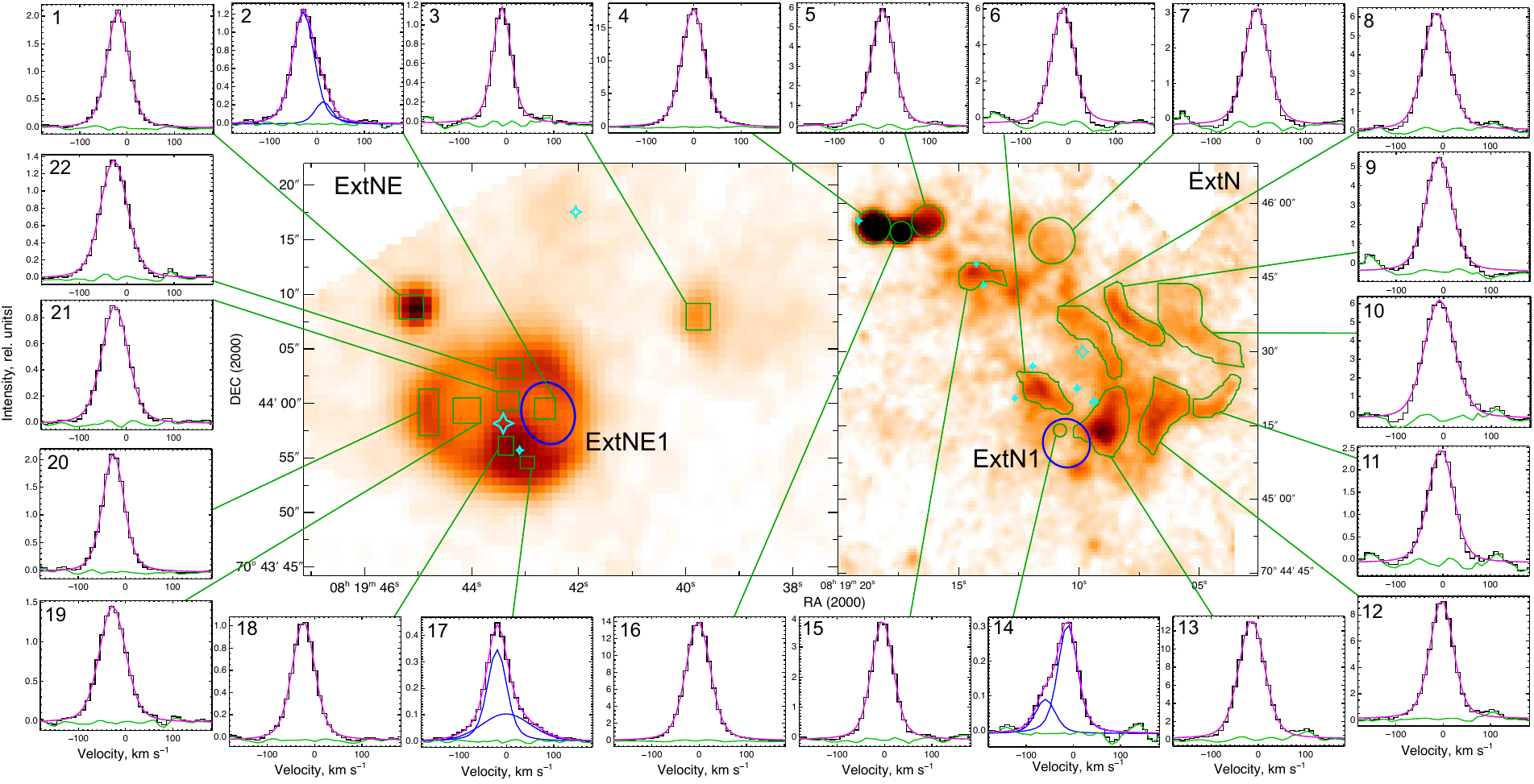}
	\caption{The same as in Fig.~\ref{fig:SE_shells} but for the ExtN (right) and ExtNE (left) complexes.}\label{fig:Ext_shells}
\end{figure*}

\begin{figure*}
	\includegraphics[width=\linewidth]{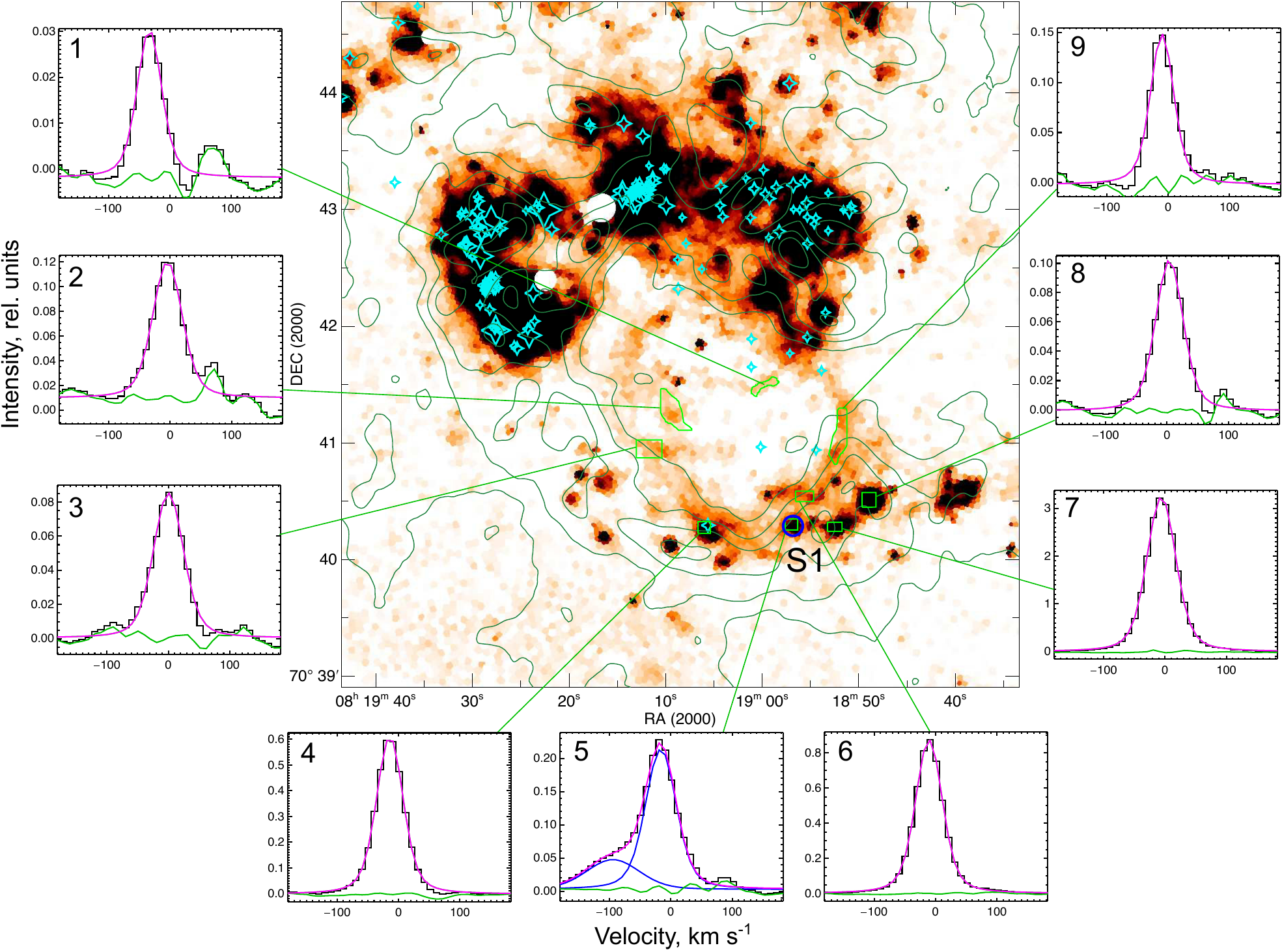}
	\caption{The same as in Fig.~\ref{fig:SE_shells} but for the whole region of ongoing star formation in the walls of the SGS. The H$\alpha$ image shown here was obtained by applying of adaptive binning with the signal-to-noise ratio equal to 30.
		Green contours show the lines of equal \HI column density.}\label{fig:sb_stars}
\end{figure*}

Deep \Ha images of the region reveal a faint filamentary structure connecting  \HII regions HSK 57, 61,
65, 67, 69 and 70 (see Fig.~\ref{fig:SE_shells}). The kinematics of neutral gas shows the existence of the local \HI superbubble that has a size of about $650\times950$~pc expands with the velocity of about $20 \kms$ and covers the whole bright part of star formation complex. The location of this \HI shell coincides with that of the \HI cavity \#27 identified by \citet{weisz09a}.

While the bright \HII regions are located in the direction of dense parts of the \HI SGS, the central part of the SE complex resides in a more exhausted medium (see the top panel of Fig.~\ref{fig:HI_maps}). Such density distribution allowed one to expect the highest expansion velocities of ionized gas in the centre of the complex. We found  3 faint expanding ionized superbubbles there which possibly interact with each other and 3 superbubbles located at the periphery of the complex (including two young compact supershells SE1 and SE6).

As is evident from the \Ha line profiles decomposition (see Fig.~\ref{fig:SE_shells}), the ionized gas kinematics is inhomogeneous in the whole SE complex. This area exhibits the most clear signs of the expansion of shell-like structures of ionized gas. For example, the northern SE2 superbubble reveals clear two-component \Ha line profile in its centre corresponding to the expansion velocity of about $37 \kms$ -- one of the highest among the ionized superbubbles we found in Ho~II. The `velocity ellipse' for SE2 -- another evidence of its expansion -- is also clearly seen in the PV diagram constructed in the direction of its elongation (\#5 in Fig.~\ref{fig:pv}).

The bright HSK~70 region containing the ULX source Holmberg~II X-1 and a
young star cluster is located in the north-eastern part of one of the inner shells. We will analyse the
kinematics of ionizing gas in its neighbourhood in our forthcoming paper \citep{egorov16}. The young cluster in that nebulae is the main source of ionizing photons in the complex. Another sources of the ionizing radiation are the star cluster in HSK~67 nebula (in particular, the brightest one among the selected O stars) at the south-east of the complex and 10 OB stars spread inside the complex. As it follows from Table~\ref{tab:ionization}, the amount of ionizing  photons emitted by the observed OB stars is consistent to that needed to ionize gas in the complex.

Note that massive OB stars are absent inside all the mentioned ionized superbubbles (except the compact one SE6) or located at their rim close to the bright nebulae. It may indicate that the source of mechanical energy needed to support the expansion of these structures can stem from leakage of wind mechanical energy from the bright \HII regions through their broken irregular shells. An order of magnitude supports such an assumption: the two OB stars inside the SE6 shell can give the wind energy leakage of $\dot E_w\sim 10^{37}\eta\ \mathrm{erg\ s}^{-1}$ ($\eta<1$ is the leakage efficiency), where the values $\dot M_w=10^{-5}~M_\odot$ yr$^{-1}$ and $v_w=2000$ km s$^{-1}$ corresponding to the high end range \citet{marko08} have been accepted for the mass-loss rate and wind velocity. On the other hand, the total radiation energy loss in the shell can be readily estimated as $\dot E_r\sim 3\times 10^{36}$ erg s$^{-1}$, while the kinetic energy consumption $\dot E_k=\dot M_{\rm sh}V_{exp}^2/2\sim 10^{36}$ erg s$^{-1}$ with $\dot M_{\rm sh}=4\pi \rho R^2V_{exp}$; here we accepted the parameters for the expanding shell SE6 as shown in Table~\ref{tab:shells_summary}, and gas density is assumed $n\sim 0.3$ cm$^{-3}$ (see estimates below); for the cooling rate, we assumed the typical value $\Lambda/(n_{\rm H}n_e) \simeq 2\times 10^{-24}$ erg cm$^3$ s$^{-1}$ at $T_e\simeq 8000$ K \citep{draine11}. It is seen that $\eta\simeq 0.4$ would be sufficient to replenish energy consumption in this shell.

As we pointed out above, the presence of the faint extended structures jointing the \HII regions  and of the \HI supershell around the whole SE star-forming complex confirms that we are dealing with a single entity here.

\subsection*{NE complex (HSK 71, 73, 74, 56 and 59)}

The filaments of ionized gas between HSK~56, 59 and 73 \HII regions as well as the brighter emission between HSK~71, 73 and 74 are clearly seen in our deep \Ha image (Fig.~\ref{fig:NE_shells}). It allows us to consider these structures as a single star formation complex.

The observed  \Ha profiles in this area are mostly single, albeit 4 faint compact expanding shell-like
structures of ionized gas can be identified (see Fig.~\ref{fig:NE_shells}). Except them, several faint shell-like structures without any signs of expansion are observed it the north part of the complex. Broadening of the \Ha profile is observed inside the ring-like  \HII region HSK~73, in whose centre we have earlier found \citep{egorov13}
an object with a rather broad \HeII 4686 line (the area with the observed \HeII emission is shown with a white circle in Fig.~\ref{fig:NE_shells}). This nebula does not reveal a clear evidence of expansion; we found signs of its expansion only at the north-western part of the region, where the young compact shell NE1 is located. The more extended shell NE2 we found at the periphery of HSK~73 region.

The complex NE is located in a dense \HI cloud on the outer side of the SGS rim. Its northern part penetrates into the adjacent \HI supershell (see Fig.~\ref{fig:colorIm}). Such distribution of the neutral gas density in the complex is consistent with the observed faint arcs of ionized gas extended to the north of bright \HII regions over more than 350~pc (see Fig.~\ref{fig:NE_shells}). The faint supershell NE3 between HSK~56 and 59 nebulae is also north elongated.

In the same way as in the SE complex,  the absence of  massive stars in the centre of all superbubbles in the NE complex (excluding NE4) indicates  the necessity of mechanical energy to escape from the bright \HII regions in order to drive it.

Massive OB stars -- the sources of required ionizing  Lyman continuum photons, are spread in the whole NE complex and do not reveal concentration to the star clusters as it was for the SE complex. The only poor star cluster is located at the south-eastern part of the complex. As it evident from   Table~\ref{tab:ionization},
   the number of identified OB stars is enough to ionize the gas in the complex.

\subsection*{N complex (HSK~45 and 41)}

We refer to the brightest \HII region in the galaxy, HSK~45, (and also adjacent compact nebula HSK~41) as an individual complex. Its \Ha images and kinematics reveal a number of interacting shell-like structures inside these \HII regions, see Fig.~\ref{fig:N_shells}.
The expansion velocity of the  \HII region HSK~45 may be as high as $35 \kms$, albeit
the observed kinematic perturbations are most likely due to the interaction between 3 identified
local shells. Based on the ionized gas kinematic analysis, we found the N2 shell to be located inside the older and more extended superbubble N1. For example, the evidence of such a structure is a more deep `dip' at the velocity ellipse in the PV diagram constructed through the both shells (see position 23~arcsec along PV diagram \#3 in Fig.~\ref{fig:pv}).

The complex is  located at the site of probable collision of three \HI supershells -- the SGS and two other to the north of it. The \HI kinematics reveals the existence of expanding \HI shell (previously unknown) that covers the whole complex. Signs of its expansion are clearly seen on PV diagram \#1 (HI) at positions of  $100-130$~arcsec along the slice in Fig.~\ref{fig:pv}.

Massive OB stars mostly  located in the centre of the HSK~45 nebula are the sources of energy for ionization and maintaining the expansion of the observed emission structures. Our deep \Ha and \SII images of the region reveal extended arcs of ionized gas in the northern part of the complex. These structures show enhanced ratios of \SIIHa, although their position on the diagnostic diagram \OIIIHb\ vs \SIIHa\ does not indicate the clear signs of collisional ionization by shocks (see Section \ref{sec:SNe} for details).
The relatively old faint shell N4 shows up among these structures. Evidence of its expansion  follows from  PV diagram \#4 in Fig.~\ref{fig:pv}. Most likely, the mechanical energy leakage from HSK~45 region should be a source of the necessary energy to drive its expansion.

Decomposition of the \Ha profile reveals the presence of a broad component inside the bright
region HSK~45 (see profiles \# 8, 9 and 19 in Fig.~\ref{fig:N_shells}). This component might be caused by the influence of the shock waves from some energetic source, e.g. the Wolf--Rayet star, to the ISM. This idea is supported by findings of \cite{hong13} who identified signs of shocks in several areas in the central part of the HSK~45 nebula from the \HST observations.


\subsection*{NW complex (HSK 15, 16, 17, 20, 25, 26)}

This is a chain of relatively compact bright \HII regions in the northern part of the complex of ongoing
star formation in the  SGS surrounded by faint shell-like structures of ionized gas with the signs of
its expansion (see Fig.~\ref{fig:NW_shells}). The bright  \HII regions exhibit a single-component \Ha profile
which is broadened in some places. Their possible expansion velocities do not exceed $11 \kms$.

The complex NW located in the direction of the dense part of the \HI SGS rim, albeit the neutral hydrogen density varies considerably there. As it follows from the bottom panel of Fig.~\ref{fig:HI_maps}, the location of the bright nebulae basically correlates with the \HI column density. Kinematics of the neutral gas reveals the existence of an \HI shell that covers the central part of the complex.  The chain of the bright \HII regions is located in the central part of that \HI shell.

According to \cite{weisz09a}, the peak of star formation was in the direction of the NW complex about 20~Myr ago. Indeed, the GALEX images reveal bright FUV emission there (see Fig.~\ref{fig:colorIm}) which poorly correlates with the distribution of the \Ha emission. As the FUV emission indicates  a recent star formation, while \Ha emission corresponds to the ongoing star formation, such disagreement in their distribution might be an evidence for the triggering of the ongoing star formation by the feedback of the recent star formation episode in the chain of compact \HII regions of the complex.

The distribution of massive OB stars does not show any concentration to star clusters. The expansion of only 2 of 5 ionized superbubbles could be explained by the stellar wind of OB stars located inside them. The mechanical energy leakage from the adjoint bright \HII regions can provide the necessary energy to drive the expansion of another 3 superbubbles.  As it follows from Table~\ref{tab:ionization}, the amount of Lyman continuum photons from the identified OB stars is enough to ionize the gas in the NW complex.

\subsection*{External complexes (ExtN and ExtNE)}

Let us analyse now the two regions marked as ExtN and ExtNE in  Fig.~\ref{fig:reg_separation} which
are not connected with the walls of the SGS discussed in this paper. As is evident from Fig.~\ref{fig:colorIm}, both these regions are located in the rims of \HI supergiant shells located next to the SGS. Note that
the ExtNE region is relatively compact and bright when observed in the \Ha line and surrounds
the compact central source of  UV radiation. Here the presence of the broad component of the \Ha
profile in the brightest part of the complex is immediately apparent, as well as the expanding shell ExtNE1 east of it (Fig~\ref{fig:Ext_shells}). Unfortunately, we cannot determine the boundaries
of this shell and, hence, its age with certainty.

The ExtN region, in turn, appears as a complex of filamentary diffuse structures surrounding compact UV
sources. It is fair to assume that \Ha radiation is caused by younger regions
in the area, where star formation is triggered by the previous generation of stars located in
accordance with the FUV emission. We found the only ionized superbubble having signs of its expansion (ExtN1) in the southern part of the complex. Generally, the \Ha profile in both northern regions can be well fitted by a single unbroadened component implying that the expansion velocity of ExtN and ExtNE
does not exceed $11 \kms$.

Despite the scarcity of massive OB stars inside both complexes, their energy input is enough to ionize the gas there. The low brightness of these \HII complexes is consistent with a small number of sources of ionizing radiation. Note that due to the above-mentioned lack of information about  stellar population of the north-western part of the complex, in Fig.~\ref{fig:Ext_shells} and in Table~\ref{tab:ionization}, we indicate only a part of OB stars which probably influence the complex.

\subsection*{Internal ionized supershell  in SGS.}

Our deep \Ha images allowed us to identify a faint diffuse component of
ionized gas  inside the SGS  close to its walls which is not
seen in the \textit{HST}/ACS and any previous images of the galaxy.
We observed an \Ha line emission inside this SGS with an average
brightness level of $3-6\times10^{-18}\ \mathrm{erg\ s^{-1}\
cm^{-2}\ arcsec^{-2}}$ and \SII line emission with an average
brightness level of about $0.5-1\times10^{-18}\ \mathrm{erg\ s^{-1}\
cm^{-2}\ arcsec^{-2}}$. The H$\alpha$ profile in the inner
ionized supershell is narrow and symmetric. Based on the difference of the line-of-sight velocities in different parts of this supershell, we may conclude that its expansion velocity does not
exceed $10-15 \kms$ which corresponds to the expansion velocity
of the neutral SGS.

Several filaments of ionized gas are observed in the central part of the area covered by internal ionized supershells. Probably, they correspond to the emission projected toward the supershell region.  \Ha line profile \#1 which shows the line-of-sight velocity shifted at almost $30~\kms$ from the mean velocity of the supershell, supports this idea.

We denote the area that corresponds to this faint supershell as Int.shell in Table~\ref{tab:ionization} and Fig.~\ref{fig:reg_separation}. Several compact \HII regions projected to the Int.shell, are located in the southern part of the \HI SGS. One of them (the S1 shell in~Table~\ref{tab:shells_summary} and  Fig.~\ref{fig:sb_stars}) shows the clear evidence of expansion.

\section{Discussion}\label{sec:discussion}

\subsection{What triggered star formation in Ho~II?}


Detailed multicolour \textit{HST} photometry of the stellar population inside each supergiant \HI shell in the Ho~II galaxy reported by \cite{weisz09a} (see also references therein) revealed stars of several age groups rather than single coeval clusters as previously believed.  It stimulated a new currently widely accepted model: neutral supergiant shells were formed from several generations of stars over several
hundreds million years. 

The main criterion of current starburst -- the H$\alpha$ emission in the neighbourhood of a newly formed
cluster of young stars -- appears for $1-2$ Myr and rapidly fades (in about 10 Myr -- the main-sequence lifetime of OB stars and the time-scale of coherent existence of young star
clusters \citealt{lada03}). The UV radiation of massive young stars of the cluster reaches its maximum at about
5  Myr, and FUV decays over much longer time-scale limiting the time period, when
the traces of star formation remain visible to  $\simeq 100 - 150$ Myr \citep{ oconnel97, stewart00}.
That is why in terms of accepted model of supergiant shells formation, the H$\alpha$ emission and 24 $\mu$m IR radiation which trace the sites of  current SF in
Irr galaxies and, in particular, in  Ho~II, do not always correlate with the location of giant \HI
supershells of  at least several hundred Myr old.

Two questions concerning star formation processed in Irr galaxies remain to be of interest
today: what actually triggers the ongoing  star formation in a particular region of a galaxy?
And how does the structure and kinematics of giant \HI cavities and shells change under the influence
of new episodes of star formation in their walls?
The analysis of the structure and kinematics of \HII regions in the walls of the SGS in the Ho~II
galaxy that we perform in this study is an attempt to shed certain light on the problem. In this section, we discuss the probable mechanism responsible for initiation of the ongoing star formation in Holmberg~II.

An analysis of deep  $H\alpha$ images of 51 Irr and diffuse galaxies allowed
\citet{hunter93} to discover a distinguishing feature of Ho~II, which consisted of the very local
disposition of \HII regions. The authors formulated the question: why there are so few
shell-like ionized structures despite the presence of many
neutral supergiant shells which most likely have been produced by the stellar wind
and SN explosions of rich OB associations?	As we pointed out above, all local star-forming complexes --- the brightest emission nebulae as well as the brightest UV and IR regions --- are located in the
northern wall of the most extended \HI cavity in the galaxy (see Fig.~\ref{fig:colorIm}). Such localization makes the question about the nature of ongoing star formation in Ho~II to be more intriguing -- why \HII regions are observed mostly in the north part of SGS, but not spread over it?

One of possible explanation of such star formation distribution might be its occurring in the densest gaseous clouds in the galaxy. Does the locations of star-forming regions in Ho~II correlate with
the gas volume density in the SGS? To answer to this question, we estimated the \HI volume density
from the LITTLE THINGS data. We constructed the 21-cm line intensity map using the technique described by
\citet{things} based on the  \HI 21~cm line intensity distribution. To convert column density into
volume density, we assumed that the
scale height of the \HI disc and its inclination to be $h=400$~pc \citep{banerjee11}
and  $i=47^\circ$ \citep{Oh11} respectively.

As a result, we establish that the distribution of  \HI volume density in the SGS discussed, where regions of ongoing star formation are located, is, on the whole, uniform with $n_\mathrm{HI}$
varying from 0.3 to $0.7\ \mathrm{cm}^{-3}$.
The inferred densities agree with the previous estimates of the density in the SGS:
$0.5\ \mathrm{cm}^{-3}$ and $0.3\ \mathrm{cm}^{-3}$ according to \citet{bagetakos11}  and \citet{puche92} respectively.

The `central arc'  of bright ionized nebulae in the Ho~II galaxy is indeed located in the
region of the highest HI column density. However, the average column density of neutral gas
at this location does not substantially exceed the density in the southern wall of the SGS which
contains small separate star-forming regions but with no extended  \Ha emission
complexes similar to the regions in the  `central arc' (see Fig.~\ref{fig:colorIm}).

Moreover, star formation in galaxies may occur in environments with sufficiently low densities.
Recent studies fail to reveal the so-called gas surface density threshold above which star formation
begins; at lower densities, the star-formation
efficiency decreases more rapidly with the decrease of gas density but does not vanish
even at  $\Sigma_{\rm HI+H_2}<0.5$ M$_\odot$ pc$^{-2}$ (see \citealt{bigiel08, bigiel10, zasov12, abram12}
and references therein) and, apparently even at lower densities (\citealt{shi14}). Summing up, we can't explain the observed location of \HII regions in Ho~II by the gas density distribution.



Our study shows that the region that separates the SGS discussed and the giant irregularly shaped
northern cavity consisting of several shells adjoining each other in the sky plane
is characterized by a very non-uniform structure and the greatest local variations
of the neutral-gas velocity in the galaxy.
The distribution of \HI velocity dispersion along the SGS discussed here shows
insignificant variations -- approximately from 7 to $16 \kms$. However,  it is interesting to verify
whether these variations are related to ongoing star formation in the SGS rims. In order to answer
this question we compared pixel-by-pixel the \HI velocity dispersion and \Ha flux that traces star
formation rate. 

\begin{figure}
    \includegraphics[width=0.97\linewidth]{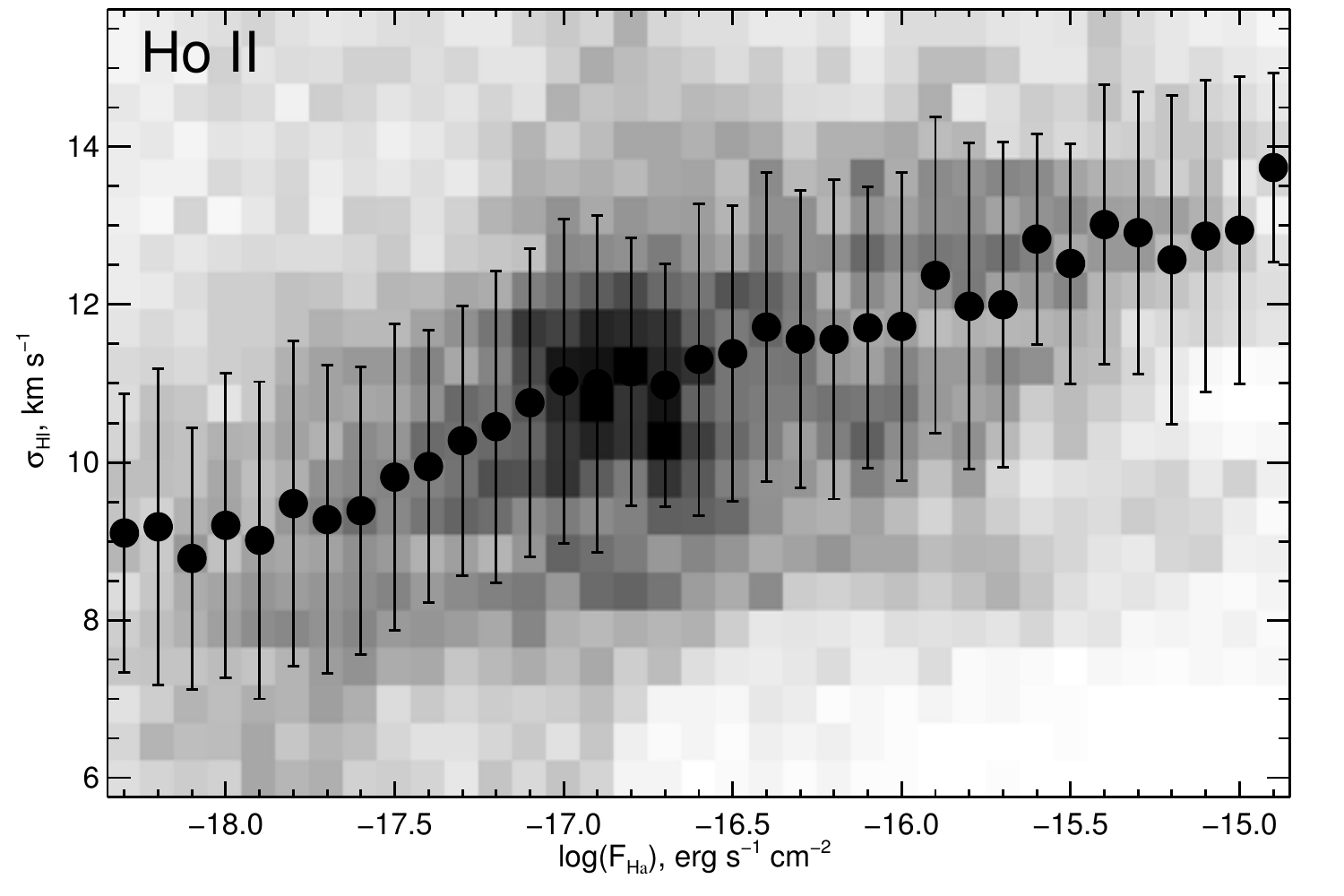}

    \includegraphics[width=0.97\linewidth]{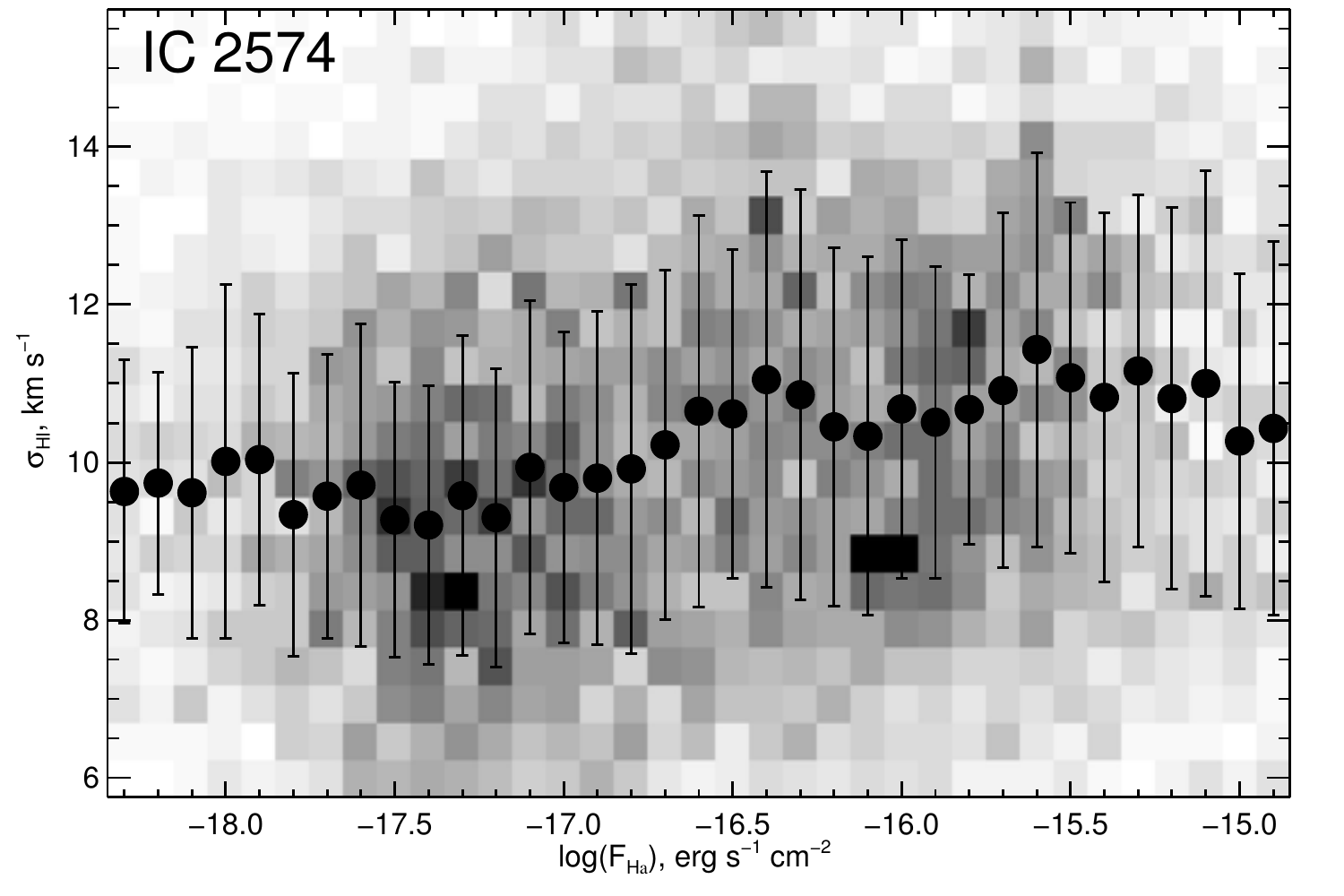}
    \caption{2D histogram of the \Ha flux and \HI velocity dispersion along the SGS in Ho~II (top) and IC~2574 (bottom) galaxies.
        Black points and bars denote the mean value and standard deviation of the \HI velocity dispersion for each \Ha flux bin.}\label{fig:HI_I-sigma}
\end{figure}

The procedure had several steps. First of all, we performed a Gaussian convolution of
our \Ha image  in order to obtain the same spatial resolution as for the \HI LITTLE THINGS data. We then resized
the convolved \Ha image in order to obtain the same pixel scale with an \HI velocity dispersion map.
After that, the pixels corresponding to the bright \HI rim of the SGS were selected for a further analysis. We
constructed a 2D histogram of the \Ha flux and \HI velocity dispersion distributions among these pixels. The
resulting histogram is shown in the top panel of Fig.~\ref{fig:HI_I-sigma}. We also show the mean \HI velocity dispersion
and its standard deviation for each flux bin.

As is evident from the constructed histogram, the \HI velocity dispersion in the SGS does not change in
the areas with a very low \Ha flux. However, starting from $F(\mathrm{H}\alpha) \simeq 2\times10^{-18} \mathrm{erg\ s^{-1}\ cm^{-2}\ arcsec^{-2}}$, velocity dispersion begins to show an increasing trend.

The observed dependence can be explained by two factors. First, the \HI velocity dispersion may have
become higher because of the collision of the SGS  with the above northern
supershells resulting in ignition of star formation in this region. Second,
the ongoing star formation itself may have an effect on the SGS by increasing the \HI turbulence
level near bright \HII regions. In the bottom panel of Fig.~\ref{fig:HI_I-sigma}, we show the similar histogram for another supergiant shell we previously studied \citep{egorov14} located in IC~2574 galaxy. In contrast to Ho~II SGS, the morphology of a supergiant shell in IC~2574 does not reveal any signs of its interaction with neighbouring shells, and we do not see any clear correlation between the \HI velocity dispersion and the \Ha flux for that case. This allows us to prefer the first explanation of the dependence observed for the Ho~II galaxy.

The above facts and considerations suggest that the last episode of star formation in the Ho~II galaxy, which
we observed as  areas of bright \HII regions and shells, was triggered by the collision of two above mentioned giant
\HI shell-like structures (or, possibly, by the collision of multiple shells that produced them).

The mechanism of star formation triggered by the collision of supergiant shells is well known
\citep[see][]{chernin95}; many colliding supergiant shells with star formation have been
found in the LMC (see the references in the above paper).

One of the most striking examples of this process is clearly observed in the irregular Local group galaxy
IC~1613, where the region of ongoing star formation is localized at the periphery of a supergiant \HI shell
at the place of its collision with another supergiant shell located north of
it \citep[see][]{lozinsk02}.

Another example of this type can be found in the NGC~6946 galaxy: a complex of \HII regions,
which is the second largest such complex in the galaxy, is located at the interface between two
neutral gas cavities \#107 and \#106 identified by \cite{boomsma08}, see also \citet*{efremov11}.

Note that the collision of supergiant shells by itself is, generally speaking, also a natural consequence
of the interaction between the collective wind and supernovae and the ambient gas over a more
extended scale length and time-scale.

\subsection{Complexes of current star formation}\label{sec:complexes}

Faint extended filamentary structures of ionized gas
identified in this study, which connect individual bright  \HII
regions, have changed our understanding of the regions of
triggered star formation in the galaxy. It is evident that
current sites of star formation are not  `chains' of bright
nebulae in the walls of the HI SGS, as previously thought, but
rather unified extended star-forming complexes with sizes
extending up to several hundreds pc,  i.e. comparable to the
size of the SGS. The conclusion that the structures considered
are single entities is further corroborated by our discovery of
faint neutral shells surrounding the complexes SE, N and  NW.

The formation of such unified complexes can be understood in terms
of the current view on the feedback between stars and the ambient
gas given the strong inhomogeneity of gas in the SGS walls. The UV
radiation of massive stars ionizes the surrounding gas and creates
bright \HII  regions. At the outer boundary of a dense parent
cloud, the ionization front penetrates further into the tenuous
medium creating  `blisters' on the side of the dense cloud (a
so-called `champagne effect') according to the model by
\cite{tenor79}. Radiation and stellar winds form faint shell-like
structures in the tenuous ambient medium. These structures
resemble  faint filamentary structures observed  in the SE, NE, N
and NW complexes. As expected in terms of this model, the observed
expansion velocity of faint shell-like formations exceeds the
expansion velocities of the bright and more compact nebulae (see
Section~\ref{sec:shells}). Note that the age of faint extended
structures should be close to that of bright and more compact
structures if these are unified sites of star formation.

Unfortunately, the accuracy of age determination for individual objects in these unified complexes of star
formation listed in  Table~\ref{tab:shells_summary} leaves a lot to be desired because
of too many uncertain factors involved.

We discussed the errors of age estimations based on the equivalent width of the H-beta line in our
earlier paper \citet{wiebe14}. The main ones are the assumption about a single starburst in a given \HII
complex and the internal extinction (see also \citealt{stasinska96}).

The age determined by \citet{stewart00} should be treated as
the mean age of stars in the aperture because of the fact that a
single-generation model is used to interpret a flux from
what is potentially a mix of populations of slightly different
ages. To compensate for this uncertainty, regions are classified into four age groups.
According to the authors, when comparing regions the actual age cutoffs for each
group should be ignored and groups should  be just thought to
consist of relatively very young, young, intermediate-age or
older star-forming regions.

In principle, the most accurate age estimate is the kinematic age if the expansion velocity
and the radius of a shell are correctly determined. However, in the case of  Ho~II
the spatial resolution is insufficient to construct
the  `velocity ellipse' in each nebula.  We attempted to estimate the expansion velocity from the
split of the position-velocity diagrams, however, no such evidence was revealed for  bright regions.
For this reason, we have estimated the expansion velocities of the
nebulae from individual profiles, i.e. without any guarantee that they refer
to the region of the most high-velocity motions. Furthermore, the irregular
structure of the  \HII regions complicates a correct determination of the
shell radius.

\begin{figure}
	\includegraphics[width=\linewidth]{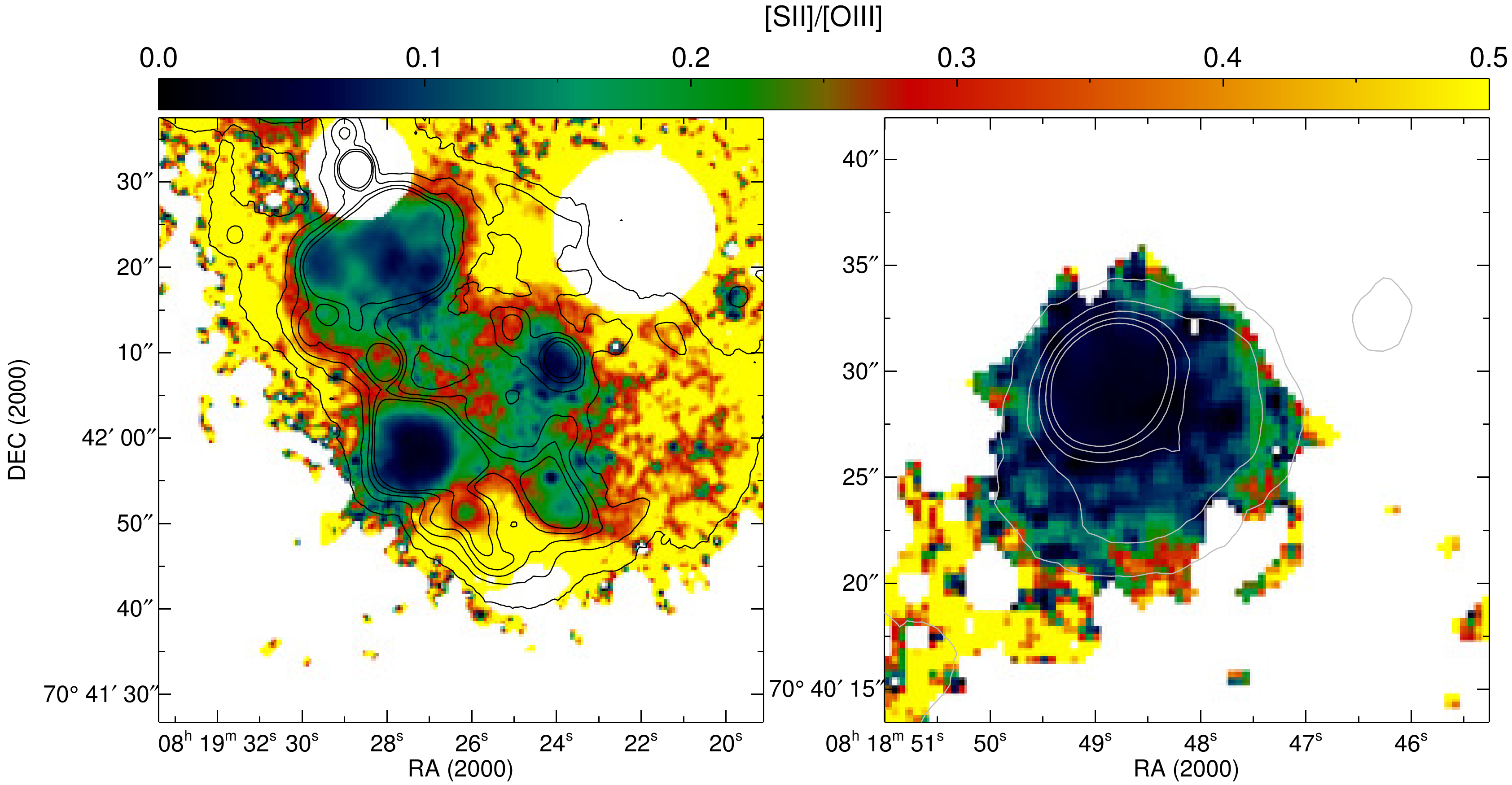}
	\caption{\SII6717,6731\AA\, to \OIII5007
		emission-line ratio maps for the  SE complex (left) and the \HII region at the south-west of the SGS rim (right). Isophotes correspond to the \Ha flux levels $(5, 15, 25, 35 and 45)\times10^{-17}\ \mathrm{erg\ s^{-1}\ cm^{-2} \ arcsec^{-2}}$. The regions contaminated by the foreground stars are masked.}\label{fig:IPM}
\end{figure}

Given these uncertainties, the ages estimated by different methods (listed
in Table~\ref{tab:shells_summary}) are consistent with our conclusion
about single complexes of  current star formation in the galaxy, which combine several
bright nebulae and the associated weak shell-like structures.

In Section~\ref{sec:morphology}, we report the estimations of the amount of ionizing radiation from
the young massive stars located inside the complexes. It follows from  Table~\ref{tab:ionization}
that it is sufficient  to account for the observed  \Ha  flux for all the complexes. In faint external complexes
the energy of stars is more than sufficient for ionizing the gas. The `extra' photons from these complexes
probably leak from these low-density and high-humidity complexes.

\cite{pellegrini11} proposed the method of distinguishing between density- and radiation bounded \HII regions based on the \SIIOIII\, ratio map. The main idea of the method is that for an optically thick nebula one should observe the increased \SIIOIII\, ratio at the edges of a region, while in the optically thin case the low ionization zone with the enhanced \SII emission might not appear.
We have applied this method to the galaxy Holmberg~II and found that almost all regions of ionized gas show the ionization structure that is typical of optically thick \HII regions -- they exhibit the outer shell of the enhanced \SIIOIII\, ratio surrounding the \OIII dominated core corresponding to the hot regions near the ionization sources. As an example of such structures, we show the \SIIOIII\, map for the SE complex on the left panel of Fig.~\ref{fig:IPM}. Note, however, the increased contribution of \OIII emission outside the border of the northern \HII region probably caused by the ionizing radiation leakage toward the centre of the SE complex. The only optically thin region in the galaxy that does not represent such a \SII shell is the compact \HII region at the south-western part of the SGS (shown in the right panel of Fig.~\ref{fig:IPM}). Thus, we may expect large escape fraction of ionizing radiation from this region, but also we cannot rule out the possible leakage from other \HII regions due to their porosity. Strictly speaking, this method allows us to identify density-bounded regions, but the regions which look like radiation-bounded in the \SIIOIII\, maps still might be optically thin.

\subsection{Internal ionized supershell in SGS.}\label{sec:superbubble}

We were the first to identify a
faint diffuse internal supershell of ionized gas  inside the SGS which was not seen in the
\textit{HST}/ACS images.  The expansion velocity of the inner shell coincides with the expansion
velocity of the neutral SGS.

The \OIIIHb\, vs \SIIHa\, diagnostic diagram shown in Fig.~\ref{fig:BPT} suggests that the
increased values of these ratios corresponding to a gas glow behind the shock front is practically not observed
in the galaxy including the region inside the the SGS.
It can be concluded from this that the nature of the ionized gas emission inside
the SGS in the Ho~II galaxy differs from that we have discovered inside the supergiant \HI shell of the IC~2574 galaxy.
Based on the observed increased \SIIHa\ and \NIIHa\ line ratios of diffuse gas in the IC~2574 SGS,
we concluded \citep{egorov14} that it should be similar to those of the extra-planar diffuse ionized
medium (DIG) in spiral and irregular galaxies.
Owing to that similarly to the DIG, we explained the faint diffuse ionized gas emission inside  the SGS
in IC~2574 as a result of leakage of both the ionizing photons and
mechanical energy from the bright \HII regions.  In the case of Ho~II, the location of the regions corresponding to the ionized internal superbubble on the diagnostic diagram (Fig.~\ref{fig:BPT}) is indicative of the photionization mechanism of gas emission there.

The analysis of the FUV morphology of the galaxy presented by \cite{stewart00} shows
the large region of faint diffuse FUV emission extending to the south-west of the central arc
of bright \HII complexes. This structure is clearly seen in Fig.~\ref{fig:colorIm}.
The relatively bright regions of  FUV emission, denoted as star-forming regions 23, 35, 36 and 37 in the list of \cite{stewart00} and selected by the above authors
as areas with FUV and no  H$\alpha$ emission and also fainter regions
practically completely fill the inner supergiant shell (see Fig.~\ref{fig:colorIm}).
However, the observed FUV emission comes from the regions, where intensive star formation occurred 20--60~Myr ago \citep{weisz09a} and hence there was a lack of ionizing Lyman quanta.  What is responsible for the creation of this internal ionized supershell in SGS then?

As it follows from Fig.~\ref{fig:sb_stars}, five identified OB stars are located inside the internal \Ha superbubble near its north-western edge. Their photometry shows that all of them are massive O stars. Correspondingly, these stars are most likely the main source of ionizing photons. Thus, according to Table~\ref{tab:ionization}, the amount of ionizing quanta from these stars is consistent with the value necessary to maintain the observed \Ha flux. Note that the pure diffuse component of the ionized superbubble is denoted as `Int.shell (diff)' in Table~\ref{tab:ionization}, while `Int.shell (all)' means the whole region including the compact bright \HII nebulae at the rim of the superbubble. Additional sources of ionizing radiation could be located in the south-western chain of bright compact \HII regions, for which there are no available data about their stellar population.

However, a certain contribution of energy leakage to the ionization of the superbubble
can not be ruled out. The latter is supported with the fact that  the brightest nebula at the southern rim of the superbubble seems to be  optically thin (see Fig.~\ref{fig:IPM} and Section~\ref{sec:complexes}) that might cause the high fraction of  ionizing photons escape.

Summing up, the ionized internal supershell was created most probably by the influence of ionizing radiation of 5 O stars located at its interior and of the additional energy leakage from nearby bright \HII complexes to the internal walls of the \HI SGS.

In addition, a continuing mechanical energy input from stellar activity, winds and SNe explosions, may also act on to the internal side of the SGS and contribute to its dynamics. For instance, speaking about the SE complex shown in Fig.~\ref{fig:SE_shells}, it can be clearly seen from rough estimates that the total mechanical energy input from the stars inside the shell $\dot E_{\rm wind}\sim 10^{38}$ erg s$^{-1}$ and the shell kinetic energy consumptions $\dot E_k\sim\sum 4\pi\rho R^2v_s^3\sim 2\times 10^{38}$ erg s$^{-1}$ are marginally equal (the gas density $n\sim 0.3$ cm$^{-3}$ is assumed, see above).

Shock waves from a stellar wind and SNe
evacuate most of gas into the shell, so that the cavity
remains filled mostly by wind and SNe ejecta with a very low density. For instance, when a supergiant shell is produced by multiple SNe explosions the remaining density might be as low as $\simlt 10^{-4}$ of the ambient density \citep{sharma14}. Even if a much less violent energy release by ionizing radiation from underlying OB stars evacuates the gas, its density within the photo-ionized bubble remains as low as $\sim 10^{-2.5}$ \citep{henney05, henney07}. One, therefore, estimates the gas density inside the bubble as $n_b\simlt 5\times 10^{-5}$
cm$^{-3}$ if the bubble is due to SNe explosions, and as $n_b\sim 10^{-3}$ cm$^{-3}$ otherwise.

In these conditions, the wind generated by massive stars inside the SGS acts from the interior and transfer momentum to the supershell. The free
expansion phase of the wind continues until the swept-up mass equals to the wind-blown mass \citep[see, e.g.][]{draine11}
\be
t_0\simeq 10^2n_b^{-1/2}\dot M_6^{1/2}v_{\rm w,8}^{-3/2}~{\rm yr},
\ee
where $\dot M_6$ in units $10^{-6}M_\odot$ yr$^{-1}$, $v_{\rm w,8}$ is the wind velocity in $10^3$ km s$^{-1}$. This gives $t_0\simeq 3\times 10^4$ yr and the corresponding radius $R_0\simeq 30$ pc for a SNe produced the SGS and an order of magnitude lower $t_0$ and $R_0$ for the
case of a SGS driven by ionization fronts.  Accounting that the massive stars inside the supershell (marked in Fig.~\ref{fig:sb_stars}) are located at larger distances -- 100 to 400 pc, one
can estimate pressure acting on to the shell as
\be \label{vs}
P_{\rm w}=\rho_b\dot R_b^2\simeq 10^{-8}\left({L_{36} n_b^2\over R_{\rm 1~pc}^2}\right)^{1/3}~{\rm dyn~cm^{-2}},
\ee
where $L_{36}$ is the wind mechanical luminosity in units $10^{36}$ erg s$^{-1}$, $R_b$ is the radius of a wind-driven shock in the ambient medium, $R_{\rm 1~pc}$ is its
value in 1 pc. When the wind-driven shock acts on to the supershell it can support the expansion with the velocity
\be
v_{_{\rm SGS}}\simeq \left({P_{\rm w}\over\rho_0}\right)^{1/2}\simeq 0.8\times 10^8{\left({L_{36}\over R_{\rm 1~pc}^2}\right)}^{1/6}{n_b^{1/3}\over n_0^{1/2}}~{\rm cm~s^{-1}},
\ee
so that for a SNe swept-up supershell with $n_b\sim 5\times 10^{-5}$ cm$^{-3}$ a star located at $R=100$ pc (as, e.g. the brightest star in the left corner of the SE5 shell in Fig.\ref{fig:SE_shells}) would  produce the velocity of the supershell of 6 km s$^{-1}$, while for $n_b=10^{-3}$ cm$^{-3}$ it would support $v{_{\rm SGS}}\simeq 13$ km s$^{-1}$. It is worth stressing that episodic wind and SNe explosions and the respective feedback during the whole evolution of the SGS can support gas density in the bubble at a level sufficient for transferring a proper momentum on its internal surface.

In those cases, when OB stars lie closer than $R_0$ to the SGS edge, they act on to the supershell by the expanding wind directly without
transferring momentum through the gas inside the cavity. In this case, the shell velocity supported by wind would be a factor of $(L_{36} n_0/R_{\rm 1~pc}^2 n_b^2)^{1/6}$ higher than
the above estimate in Eq.~(\ref{vs}). One may expect that the similar enhancement of the action of a stellar wind on the supershell would take place, when the wind propagates through
low density tunnels inside the bubble. Overall, one may think that energy release in the form of Ly-continuum photons and corpuscular winds emitted by massive stars inside the supershells does not only support their expansion, but also provides proper conditions for transferring momentum from stars to the walls and keeping them continuously expanding until star formation is exhausted.

\subsection{Search for SNRs in the SGS.}\label{sec:SNe}

Radio observations of the Ho~II galaxy  \citep*{tong95, braun07, heald09} revealed a continuum radio emission
in the region of bright emission nebulae.
The synchrotron component
of radio emission was identified in the eastern chain of bright nebulae by \citet{tong95}
and this fact led the authors to suspect that these areas may contain supernova remnants.
The detected polarized radio emission also coincides with this chain \citep{heald09}.


\citet{hong13} identified the shock-ionized component via the line diagnostic
diagram \OIIIHb\ vs \NIIHa\ inside the HSK~45 nebula using the high resolution \HST data.

Our spectroscopic observations \citep{egorov13} did not allow us
to identify the optical emission of these hypothetical supernova
remnants and shock fronts by the \textsc{I([S~ii])/I(H$\alpha$)}
line ratio.

\begin{figure}
    \includegraphics[width=\linewidth]{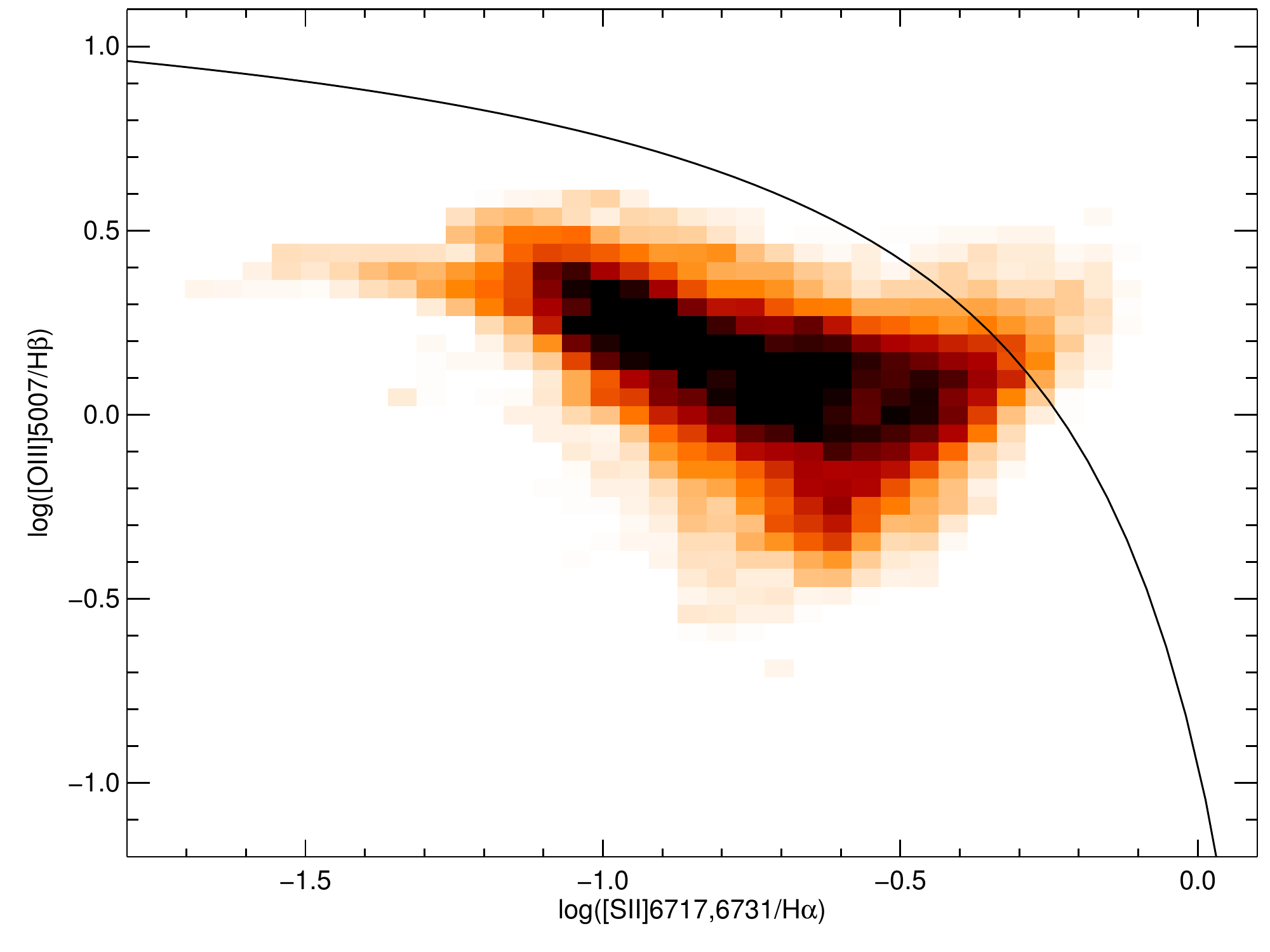}
    \caption{Diagnostic diagram $\log(\textsc{[O iii]}5007/\mathrm{H}\beta)$ (computed by dividing the \Ha flux by 2.86) vs $\log(\textsc{[S ii]}6717,6731/\mathrm{H}\alpha)$ constructed for our narrow-band images in H$\alpha$, \SII and \OIII lines.  A separation line from \citet{kewley01} between regions of pure photoionization excitation and of significant contribution of the shocks is shown by a black line.}\label{fig:BPT}
\end{figure}

We used our narrow-band images in the \OIII 5007\AA, \SII 6717,6731\AA\, and \Ha lines to construct the diagnostic diagram of \OIIIHb\, vs \SIIHa\, ratios for each pixel of the galaxy. All the used line ratios were corrected by reddening using the mean value $E(B-V)=0.05$ \citep{egorov13}. We calculated the values of H$\beta$ fluxes  from the H$\alpha$ flux using the theoretical ratio I(H$\alpha$)/I(H$\beta$) for $T=10000$~K \citep{osterbrock}. The result is shown in Fig.~\ref{fig:BPT}. A black curve denotes the separation line between the regions with pure photoionization and composite (with a probably significant contribution of the shocks) excitation \citep{kewley01}.  As  follows from this diagram, almost all emission observed in the galaxy has been excited by photoionization and do not show any signs of shocks. The points lying over the separation curve in the diagram correspond to the regions of low brightness with low signal-to-noise ratio. Nevertheless, we should note a limited implication of this method of construction of these diagnostic diagrams, because the inhomogeneous reddening may lead to incorrect results. Also the low metallicity should shift the separation line to the area of lower line ratios. But even proposing that all points corresponding to the right wing of the `seagull shape' in Fig.~\ref{fig:BPT} should lie under the curve, the collision excitation revealed by that way will be important only in the diffuse surrounding of the HSK~45 nebula.

If shock waves from possible SNe and/or stellar winds play an important role in the excitation of emission
lines in some regions, we should be able to reveal the corresponding kinematic signatures.
In bright nebulae of the eastern chain, where radio observations suggested the presence of
supernova remnants, we found no signs of high velocities typical of not too old supernova remnants.
However, high-velocity components of the \Ha line (in some places having the elevated velocity dispersion) are indeed observed in the outer weakly ionized
structures in the SE and NE areas and especially in the N area, where the brightest
nebula,  HSK 45, is located (see Figs.~\ref{fig:SE_shells}, \ref{fig:NE_shells} and \ref{fig:N_shells}).
We currently can not say firmly whether these high-velocity motions are
associated with the inflow of kinetic energy from supernova explosions or from the winds of O type stars.  Note, however, that when shock waves from supernovae propagate in an inhomogeneous (cloudy) medium,
all typical shock manifestations --  via emission, morphology or kinematics,  become amorphous \citep{korolev15}.
Hence, given that the region under study has been a subject of repeated exposure
of strong perturbing factors (ionization fronts, wind flows) over long time, the absence of clear signs
of supernovae can be explained by the inhomogeneity of the gaseous environment.

\section{Summary}\label{sec:summary}

A detailed analysis of the structure and kinematics of all bright
\HII complexes of ongoing star formation in the walls of the
supergiant shell of neutral gas in the Ho~II galaxy is performed
based on the observations carried out at the SAO RAS 6-m telescope
with a scanning Fabry--Perot interferometer  in the \Ha line
combined with direct images taken in the H$\alpha$, \SII and \OIII
lines. We also used the {\it HST} archival images. The kinematics of
ionized gas is compared to that of neutral gas based on the data
of VLA observations in the 21-cm radio line taken from the LITTLE THINGS survey archive \citep{littlethings} and to the stellar population
of the area.

The observed data  sheds certain light on the process of evolution
of giant \HI structures under the action by ionizing radiation and inflow of mechanical energy
from local bursts of star formation in the walls of these structures.

The following results have been obtained:

\begin{enumerate}

    \item We found 22 faint expanding ionized superbubbles in star formation complexes of the galaxy and estimated their
    expansion velocities and kinematic ages using the results of the ionized gas kinematics analysis. Also 3 local expanding \HI shells tied with star formation complexes in the SGS rim were identified.

    \item 
We showed that current star formation episodes in the SGS  were
more extensive  and complex than previously thought: they
represent not  `chains' of separate individual bright nebulae in
the walls of the \HI SGS but rather unified star-forming
complexes with sizes of several hundred pc.
The formation of such
unified complexes is due to the  stellar feedback given the
strong inhomogeneity of the gas in the SGS walls.
Given the large errors of different methods, the inferred age of
faint extended and bright more compact structures is consistent
with the assumption that they are unified complexes of ongoing
star formation.

\item We suggest that the last episode of star formation in the galaxy that we observe as  areas of bright \HII regions and shells
in the northern wall of the  SGS was triggered by its collision with giant shell-like  \HI
structures located north of the SGS.

    \item We discovered a faint ionized supershell inside the neutral SGS. 
    The origin of this weak  \Ha emission somewhat differs from that of
    the faint inner ionized supershell that we earlier found in
    the SGS in IC~2574 galaxy, where  the  leakage of ionizing photons from bright \HII regions in the walls of the SGS is
     proposed as a main  ionization source \citep{egorov14}.  In Ho~II,  five OB stars located inside
    the ionized supershell can explain its emission; however,   we do not also
    rule out the  leakage of ionizing photons from bright \HII regions.

    \item  We have not been found any clear kinematic signatures
    of the effect of shock waves associated
    with supernova remnants in the eastern chain of bright  nebulae earlier
    suspected of the synchrotron component of the radio emission.

\end{enumerate}


\section*{Acknowledgements}

We are thankful to the anonymous referee for the constructive comments.

This work is based on observations obtained with the 6-m telescope of the
Special Astrophysical Observatory of the Russian Academy of Sciences carried out with the financial
support of the Ministry of Education and Science of the Russian Federation
(agreement No. 14.619.21.0004, project ID RFMEFI61914X0004).

The study  was supported by the Russian
Foundation for Basic Research (projects 14-02-00027, 15-32-21062, 15-02-08293 and 15-52-45114-IND). The study of gas kinematics was supported by the RSCF grant No. 14-22-00041. YS is also thankful to
the Grant of the President of the Russian Federation for Support of the Leading Scientific Schools NSh-4235.2014.2. AVM is also grateful   for the financial support via the grant
MD3623.2015.2 from the President of the Russian Federation.

This research has made use of
the NASA/IPAC Extragalactic Database (NED) which is operated by the Jet
Propulsion Laboratory, California Institute of Technology, under contract with
the National Aeronautics and Space Administration, and of the Lyon Extragalactic
Database (LEDA). 

	The study is partially based on observations made with the NASA/ESA Hubble Space Telescope, obtained from the Data Archive at the Space Telescope Science Institute, which is operated by the Association of Universities for Research in Astronomy, Inc., under NASA contract NAS 5-26555. These observations are associated with programs \# 10522 and \# 10605.

\label{lastpage}


\begin{thebibliography}{101}

    \bibitem[\protect\citeauthoryear{Abramova \& Zasov}{2012}]{abram12}
    Abramova O. V., Zasov  A. V., 2012,
    Astron.Letters, 38, 755


    \bibitem[\protect\citeauthoryear{Afanasiev \& Moiseev}{2011}]{scorpio2}
    Afanasiev V.L., Moiseev A.V., 2011, Baltic Astronomy, 20, 363




    \bibitem[\protect\citeauthoryear{Bagetakos et al.}{2011}]{bagetakos11}
    Bagetakos I., Brinks E., Walter F., de Blok W.J.G., Usero~A., Leroy A.K., Rich J.W., Kennicutt R.C., 2011, AJ, 141, 2011


    \bibitem[\protect\citeauthoryear{Banerjee et al.}{2011}]{banerjee11}
    Banerjee A., Jog C.J., Brinks E., Bagetakos I., 2011, MNRAS, 415, 687


    \bibitem[\protect\citeauthoryear{Bastian et al.}{2011}]{bastian11}
    Bastian N. et al., 2011, MNRAS, 412, 1539

%
%

    \bibitem[\protect\citeauthoryear{Bernard et al.}{2012}]{bernard12}
    Bernard E.J., Ferguson A. M. N.,  Barker M.K., Irwin  M.J., Jablonka P.,
    Arimoto N., 2012, MNRAS, 426, 3490


    \bibitem[\protect\citeauthoryear{Bigiel et al.}{2008}]{bigiel08}
    Bigiel F., Leroy A., Walter F., Brinks E., de Blok W. J. G., Madore
    B., Thornley M. D., 2008, AJ, 136, 2846

    \bibitem[\protect\citeauthoryear{Bigiel et al.}{2010}]{bigiel10}
    Bigiel F., Leroy A., Walter F., Blitz L., Brinks E., de Blok W. J. G.,  Madore
    B. , 2010, AJ, 140, 1194

    \bibitem[\protect\citeauthoryear{Boomsma et al.}{2008}]{boomsma08}
    Boomsma R., Oosterloo T.A., Fraternali F., van der Hulst J., Sancisi R. , 2008, A\&A, 490, 555


    \bibitem[\protect\citeauthoryear{Braun et al.}{2007}]{braun07}
    Braun, R., Oosterloo, T. A., Morganti, R., Klein, U., Beck, R., 2007, AAp, 461, 455

    \bibitem[\protect\citeauthoryear{Bureau \& Carignan}{2002}]{bureau02}
    Bureau M.,  Carignan C., 2002, AJ, 123, 1316


    \bibitem[\protect\citeauthoryear{Camps-Farin{\~a} et al.}{2015}]{cf15}
Camps-Farin{\~a} A., Zaragoza-Cardiel J., Beckman J.E. , Font J., Garc{\'i}a-Lorenzo B.,  Erroz-Ferrer S., Amram P., 2015, MNRAS, 447, 3840

    \bibitem[\protect\citeauthoryear{Cannon et al.}{2011a}]{cannon11a}
    Cannon  J. M. et al. 2011a, ApJ, 735, 35C

    \bibitem[\protect\citeauthoryear{Cannon et al.}{2011b}]{cannon11b}
    Cannon  J. M. et al. 2011b, ApJ, 735, 36C

    \bibitem[\protect\citeauthoryear{Chernin, Efremov \& Voinovich}{Chernin et al.}{1995}]{chernin95}
    Chernin A.D., Efremov, Yu.N., Voinovich, P.A., 1995, MNRAS, 275, 313

   \bibitem[\protect\citeauthoryear{Chevalier}{1974}]{chevalier74}
	Chevalier R.A., 1974, ApJ, 188, 501

 \bibitem[\protect\citeauthoryear{Cox}{1972}]{cox72}
 Cox D. P., 1972, ApJ, 178, 159
%
%
%
%
%
%

    \bibitem[\protect\citeauthoryear{Dalcanton et al.}{2008}]{angst}
    Dalcanton J.J. et al., 2008 ApJS, 183, 67



    \bibitem[\protect\citeauthoryear{Dicaire et al.}{2008}]{dicaire08}
    Dicaire I. et al., 2008 MNRAS, 385, 553


   \bibitem[\protect\citeauthoryear{Dib \& Burkert}{2005}]{dib05}
   Dib S., Burkert A. 2005, ApJ, 630, 238

    \bibitem[\protect\citeauthoryear{Draine}{2011}]{draine11}
    Draine, B. T., 2011, Physics of  the Interstellar and Intergalactic Meduim. Princeton


    \bibitem[\protect\citeauthoryear{Efremov, Afanasiev \& Egorov}{Efremov et al.}{2011}]{efremov11}
    Efremov, Yu.N., Afanasiev V.L., Egorov O.V., 2011, Astron.Bull., 66, 304

	\bibitem[\protect\citeauthoryear{Egorov, Lozinskaya \& Moiseev}{Egorov et al.}{2010}]{egorov10}
	Egorov O.V., Lozinskaya T.A., Moiseev A.V., 2010, Astron.Rep., 54, 277

    \bibitem[\protect\citeauthoryear{Egorov, Lozinskaya \& Moiseev}{Egorov et al.}{2013}]{egorov13}
    Egorov O.V., Lozinskaya T.A., Moiseev A.V., 2013, MNRAS, 429, 1450

    \bibitem[\protect\citeauthoryear{Egorov et al.}{2014}]{egorov14}
    Egorov O.V., Lozinskaya T.A., Moiseev A.V., Smirnov-Pinchukov G.V., 2014, MNRAS, 444, 376

    \bibitem[\protect\citeauthoryear{Egorov, Lozinskaya \& Moiseev}{Egorov et al.}{2015}]{egorov15}
    Egorov O.V., Lozinskaya T.A., Moiseev A.V., 2015, Astron.Astrophys.Trans., 29, 17

\bibitem[\protect\citeauthoryear{Egorov, Lozinskaya \& Moiseev}{Egorov et al.}{2016}]{egorov16}
Egorov O.V., Lozinskaya T.A., Moiseev A.V., 2016, MNRAS, submitted

   \bibitem[\protect\citeauthoryear{Elmegreen}{1997}]{elmegreen97}
  Elmegreen B.~G. 1997, ApJ, 477, 196

   \bibitem[\protect\citeauthoryear{Elmegreen \& Chiang}{1982}]{elmegreen82}
   Elmegreen B.~G.,  Chiang, W.-H., 1982, ApJ, 253, 666

%
%
%
%
%


    \bibitem[\protect\citeauthoryear{Heald, Braun \& Edmonds}{Heald et al.}{2009}]{heald09}
    Heald G., Braun R., Edmonds R., 2009, A\&A, 503, 409


    \bibitem[\protect\citeauthoryear{Henney et al.}{2005}]{henney05}
    Henney, W. J., Arthur, S. J., Williams, R. J. R., Ferland, G. J., 2005, ApJ, 621, 328

    \bibitem[\protect\citeauthoryear{Henney}{2007}]{henney07}
    Henney, 2007, in Hartquist T.W., Pittard J. M., Falle. S. A. E. G. eds., Astrophys. \& Space Sci. Proc., 1,  Diffuse Matter from Star Forming Regions to Active Galaxies, Springer, Dordrecht, p.103


    \bibitem[\protect\citeauthoryear{Hodge, Strobel \& Kennicutt}{Hodge et al.}{1994}]{hodge94}Hodge P., Strobel N.V., Kennicutt R.C., PASP, 1994, 106, 309

    \bibitem[\protect\citeauthoryear{Hong et al}{2013}]{hong13}
    Hong S., Calzetti D., Gallagher J.S., Martin C.L., Conselice C.J., Pellerin A., 2013, ApJ, 777, 63


    \bibitem[\protect\citeauthoryear{Hunter, Hawley \& Gallagher}{Hunter et al.}{1993}]{hunter93}
    Hunter D.A., Hawley W.N., Gallagher J.S., 1993, AJ, 106, 1797

    \bibitem[\protect\citeauthoryear{Hunter et al.}{2012}]{littlethings}
    Hunter D.A. et al., 2012, AJ, 144, 134

    \bibitem[\protect\citeauthoryear{Hunter, Elmegreen \& Gehret}{Hunter et al.}{2016}]{hunter16}
    Hunter D.A., Elmegreen B.G., Gehret E., 2016, AJ, 151, 136

%

    \bibitem[\protect\citeauthoryear{Karachentsev \& Kaisin}{2007}]{karach07}
    Karachentsev I.D.,  Kaisin S.S., 2007, AJ, 133, 1883

    \bibitem[\protect\citeauthoryear{Karachentsev, Makarov \& Kaisina}{Karachentsev et al.}{2013}]{karach13}
    Karachentsev I.D., Makarov D.I., Kaisina E.I., 2013, AJ, 145, 101


    \bibitem[\protect\citeauthoryear{Kennicutt et al.}{2003}]{sings}
    Kennicutt R.C. Jr. et al., 2003, PASP, 115, 928

    \bibitem[\protect\citeauthoryear{Kennicutt et al.}{2011}]{kingfish}
    Kennicutt R.C. et al., 2011, PASP, 123, 1347

 \bibitem[\protect\citeauthoryear{Kewley et al.}{2001}]{kewley01}
Kewley L. J., Dopita M. A., Sutherland R. S., Heisler C. A., Trevena J., 2001, ApJ, 556, 121

   \bibitem[\protect\citeauthoryear{Kim et al.}{1999}]{kim99}
   Kim S., Dopita M. A., Stavelet-Smith L., Bessel M., 1999, A\&A, 350, 230
%

    \bibitem[\protect\citeauthoryear{Korolev et al.}{2015}]{korolev15}
    Korolev V.V., Vasiliev E.O., Kovalenko I.G., Shchekinov Yu.A., 2015, ARep, 59, 690

    \bibitem[\protect\citeauthoryear{Lada \& Lada}{2003}]{lada03}
    Lada C.J., Lada E.A., ARA\&A, 41, 57, 2003.


    \bibitem[\protect\citeauthoryear{Lehmann et al.}{2005}]{lehmann05}
    Lehmann I et al., 2005, A\&A, 431, 847

    \bibitem[\protect\citeauthoryear{Leitherer et al.}{1999}]{leitherer99}
    Leitherer C. et al., 1999, ApJ, 123, 3

    \bibitem[\protect\citeauthoryear{Leroy et al.}{2009}]{heracles}
    Leroy A.K. et al., 2009, AJ, 137, 4670

%
%
%

    \bibitem[\protect\citeauthoryear{Lozinskaya}{2002}]{lozinsk02}
    Lozinskaya T.A., 2002, Astron. Astroph. Transactions, 21, 223

 \bibitem[\protect\citeauthoryear{Lozinskaya, Moiseev \& Podorvanyuk}{Lozinskaya et al.}{2003}]{lozinsk03}
	Lozinskaya T.A., Moiseev A.V., Podorvanyuk N.Yu., 2003, RMXAA, 15, 284



    \bibitem[\protect\citeauthoryear{Markova \& Puls}{Markova \& Puls}{2008}]{marko08}
     Markova N. \& Puls J., 2008, A\&A, 478, 823

    \bibitem[\protect\citeauthoryear{Martins, Schaerer \& Hillier}{Martins et al.}{2006}]{Martins05}
    Martins F., Schaerer D., Hilliier D.J., 2005, A\&A, 436, 1049

    \bibitem[\protect\citeauthoryear{Maschenko \& Silich}{1995}]{maschenko95}
    Maschenko S.Ya., Silich S.A., 1995, Astron.Rep., 39, 587

%
   \bibitem[\protect\citeauthoryear{McCray \& Kafatos}{1987}]{mccray87}
   McCray R., Kafatos~M., 1987, ApJ, 317, 190


    \bibitem[\protect\citeauthoryear{McQuinn et al.}{2009}]{mcquinn09}
    McQuinn K.B.W., Skillman E.D., Cannon J.M., \mbox{Dalcanton}~J.~J., Dolphin A., Stark D., Weisz D., 2009, ApJ, 695, 561

    \bibitem[\protect\citeauthoryear{McQuinn et al.}{2010a}]{mcquinn10a}
    McQuinn K.B.W. et al., 2010a, ApJ, 721, 297

    \bibitem[\protect\citeauthoryear{McQuinn et al.}{2010b}]{mcquinn10b}
    McQuinn K.B.W. et al., 2010b, ApJ, 724, 49

%

    \bibitem[\protect\citeauthoryear{Moiseev}{2002}]{Moiseev02ifp}
    Moiseev A.V., 2002, Bull. Spec. Astrophys. Obs., 54, 74

    \bibitem[\protect\citeauthoryear{Moiseev \& Egorov}{2008}]{MoiseevEgorov2008}
    Moiseev A.V., Egorov O.V., 2008, Astrophysical Bulletin, 63, 181

    \bibitem[\protect\citeauthoryear{Moiseev \& Lozinskaya}{2012}]{ML12}
    Moiseev A.V., Lozinskaya T.A., 2012, MNRAS, 423, 1831


    \bibitem[\protect\citeauthoryear{Moiseev}{2014}]{Moiseev2014}
    Moiseev A.V., 2014, Astrophysical Bulletin, 69, 1

    \bibitem[\protect\citeauthoryear{Moiseev}{2015}]{Moiseev2015}
    Moiseev A.V., 2015, Astrophysical Bulletin, 70, 494



    \bibitem[\protect\citeauthoryear{Moustakas et al.}{2010}]{moustakas10}
    Moustakas J., Kennicutt R.C. Jr., Tremonti C.A., Dale D.A., Smith J.-D. T., Calzetti  D., 2010, ApJS, 190, 233

    \bibitem[\protect\citeauthoryear{Mu\~{n}oz-Tu\~{n}\'{o}n et al.}{1996}]{MunozTunon1996}
Mu\~{n}oz-Tu\~{n}\'{o}n C., Tenorio-Tagle G., Casta\~{n}eda H.O., Terlevich R., 1996, AJ, 112, 1636

    \bibitem[\protect\citeauthoryear{Nath \& Shchekinov}{2013}]{nath13}
    Nath, B.B. \& Shchekinov, Yu., 2013, ApJL, 777, L12


    \bibitem[\protect\citeauthoryear{Oh et al.}{2011}]{Oh11}
    Oh S.-H., de Blok W. J. G.,  Brinks E., Walter F., \mbox{Kennicutt}~R.~C.~Jr., 2011, AJ, 141, 193

\bibitem[\protect\citeauthoryear{Ott et al.}{2001}]{ott01}
Ott J., Walter F., Brinks E., Van Dyk S. D., Dirsch B., Klein U., 2001, AJ, 122, 3070

    \bibitem[\protect\citeauthoryear{O'Connell}{1997}]{oconnel97}
    O'Connell R.W., 1997, AIP Conf. Proc., 408, 11

    \bibitem[\protect\citeauthoryear{Osterbrock \& Ferland}{2006}]{osterbrock}
    Osterbrock D.E, Ferland G.J., 2006, Astrophysics of Gaseous Nebulae and Active Galactic Nuclei, 2nd Edition.  Univ. Sci. Books



%
%
%
%
%

  \bibitem[\protect\citeauthoryear{Pellegrini et al.}{2011}]{pellegrini11}
  Pellegrini E.W., Oey M.S, Winkler P.F., Smith R.C., Points S.,  2011, Bull. Soc. R. Sci. de Li\'ege, 80, 410


    \bibitem[\protect\citeauthoryear{Puche et al.}{1992}]{puche92}
    Puche D., Westpfahl D., Brinks E., Roy J.-R., 1992, AJ, 103, 184

	\bibitem[\protect\citeauthoryear{Rela{\~n}o \& Beckman}{2005}]{relano05}
	Rela{\~n}o M., Beckman J. E., 2005, A\&A, 430, 911

    \bibitem[\protect\citeauthoryear{Rhode et al.}{1999}]{rhode99}
    Rhode K. L., Salzer J. J., Westpfahl D., Radice L. A. 1999, AJ, 118, 323


    \bibitem[\protect\citeauthoryear{S{\'a}nchez-Salcedo, Hidalgo-G{\'a}mez \& Mart{\'i}nez-Garc{\'i}a}{S{\'a}nchez-Salcedo et al.}{2014}]{HoII_incl_new}
    S{\'a}nchez-Salcedo F.J., Hidalgo-G{\'a}mez A.M., Mart{\'i}nez-Garc{\'i}a E.E., 2014, Rev.Mex, 50, 225




%


	\bibitem[\protect\citeauthoryear{S\'anchez-Cruces et al.}{2015}]{sc15}
	S{\'a}nchez-Cruces M., Rosado M., Rodr{\'i}guez-Gonz{\'a}lez A., Reyes-Iturbide J., 2015, ApJ, 799, 231

    \bibitem[\protect\citeauthoryear{Schaerer \& Vacca}{1998}]{schaerer98}
    Schaerer D., Vacca W. D., 1998, ApJ, 497, 618


    \bibitem[\protect\citeauthoryear{Sharma et al.}{2014}]{sharma14}
    Sharma P., Roy A., Nath B.B., Shchekinov Yu., 2014, MNRAS, 443, 3463

    \bibitem[\protect\citeauthoryear{Shi et al.}{2014}]{shi14}
    Shi Y., Armus L. Helou G., Stierwalt S., Gao Y. Wang J., Zhang Z., Gu Q., 2014, Nature, 514, 335


\bibitem[\protect\citeauthoryear{Simpson, Hunter \& Knezek}{Simpson at al.}{2005}]{Simpson05}   Simpson C.E., Hunter D.A., Knezek P.M., 2005, AJ, 129, 160


   \bibitem[\protect\citeauthoryear{Silich et al.}{2006}]{silich06}
   Silich S., Lozinskaya T., Moiseev A., Podorvanyuk N., Rosado M., Borissova J., Valdez-Guti\'errez M., 2006, A\&A, 448, 123
%
%

    \bibitem[\protect\citeauthoryear{Stasi\'nska \& Leitherer}{1996}]{stasinska96}
    Stasi\'nska G., Leitherer C., 1996, ApJS, 107, 661

    \bibitem[\protect\citeauthoryear{Stewart et al.}{2000}]{stewart00}
    Stewart S.G. et al., 2000, ApJ, 529, 201

    \bibitem[\protect\citeauthoryear{Tenorio-Tagle}{1979}]{tenor79}
    Tenorio-Tagle G., 1979, A\&A, 71, 59

\bibitem[\protect\citeauthoryear{Tenorio-Tagle}{1981}]{tenor81}
Tenorio-Tagle~G., 1981, A\&A, 94, 338

\bibitem[\protect\citeauthoryear{Tenorio-Tagle \& Bodenheimer}{1988}]{tenor88}
Tenorio-Tagle G., Bodenheimer P. 1988, ARA\&A, 26, 145

%

    \bibitem[\protect\citeauthoryear{Tomisaka, Habe \& Ikeuchi}{1981}]{tomisaka81}
    Tomisaka, K., Habe, A., Ikeuchi, S., 1981, Ap\&SS, 78, 273

    \bibitem[\protect\citeauthoryear{Tomita et al.}{1998}]{tomita98}
    Tomita A., Ohta K., Nakanishi K., Takeuchi T.T., Saito M., 1998, AJ, 116, 131

    \bibitem[\protect\citeauthoryear{Tongue \& Westpfahl}{1995}]{tong95}
    Tongue T.D., Westpfahl D.J., 1995, AJ, 109, 2462

     \bibitem[\protect\citeauthoryear{Vasiliev, Nath \& Shchekinov}{2015}]{vasiliev15}
        Vasiliev E. O., Nath, B. B., Shchekinov, Yu. A., 2015, MNRAS, 446, 1703

   \bibitem[\protect\citeauthoryear{Vorobyov \& Shchekinov}{2004}]{vorobyov04}
   Vorobyov E.~I., Shchekinov Y.~A. 2004, A\&A, 416, 499

   \bibitem[\protect\citeauthoryear{Wada, Spaans \& Kim}{Wada et al.}{2000}]{wada00}
   Wada K., Spaans M., Kim S., 2000, ApJ, 540, 797
%
%

    \bibitem[\protect\citeauthoryear{Walter et al.}{2007}]{walter07}
    Walter  F., 2007, ApJ, 661, 102

    \bibitem[\protect\citeauthoryear{Walter et al.}{2008}]{things}
    Walter F., Brinks E., de Blok W. J. G., Bigiel F., \mbox{Kennicutt}~R.~C.~Jr, Thornley M.D., Leroy A.,  2008, AJ, 136, 2563

    \bibitem[\protect\citeauthoryear{Warren et al.}{2011}]{Warren11}
    Warren S.R. et al., 2011, ApJ, 738, 10

    \bibitem[\protect\citeauthoryear{Weaver et al.}{1977}]{weaver77}
    Weaver R., McCray R., Castor J., Shapiro P., Moore R., 1977, ApJ, 218, 377

    \bibitem[\protect\citeauthoryear{Weisz et al.}{2008}]{weisz08}
    Weisz D. R., Skillman E. D., Cannon J. M., Dolphin A. E., Kennicutt R. C., Lee J., Walter F., 2008, ApJ, 689, 160


    \bibitem[\protect\citeauthoryear{Weisz et al.}{2009a}]{weisz09a}
    Weisz D. R., Skillman E. D., Cannon J. M., Dolphin A. E., Kennicutt R. C., Lee J., Walter F. 2009a, ApJ, 704, 1538

    \bibitem[\protect\citeauthoryear{Weisz et al.}{2009b}]{weisz09b}
    Weisz D.R., Skillman E.D., Cannon J.M., Walter F., Brinks~E., Ott J., Dolphin A.E., 2009b, ApJ, 691, L59

    \bibitem[\protect\citeauthoryear{Wiebe et al.}{2014}]{wiebe14}
    Wiebe D.S., Khramtsova M.S., Egorov O.V., Lozinskaya T.A., 2014, Astron. Letters, 40, 278

%
%
%

\bibitem[\protect\citeauthoryear{Yadav et al.}{2016}]{yadav16}
Yadav~N. Mukherjee~D.,  Sharma~P., Nath~ B. arxiv:1603.00815

\bibitem[\protect\citeauthoryear{Young \& Lo}{1997}]{Young97}
Young~L. M., Lo~K. Y. 1997, ApJ, 490, 710

    \bibitem[\protect\citeauthoryear{Zasov \& Abramova}{2012}]{zasov12}
    Zasov  A. V., Abramova O. V., Astron Astroph Trans, 2012, 27, 351

\end{thebibliography}
\end{document}